\documentclass[12pt,preprint]{aastex}

\def\fluxu{\,{\rm erg} \ {\rm cm}^{-2} \ {\rm s}^{-1}}

\def\lsim{\mathrel{\hbox{\rlap{\lower.55ex \hbox {$\sim$}}\kern-.0em 
\raise.4ex \hbox{$<$}}}}  
\def\gsim{\mathrel{\hbox{\rlap{\lower.55ex \hbox {$\sim$}}\kern-.0em 
\raise.4ex \hbox{$>$}}}}

\def\arcmin{\hbox{$^\prime$}}

\def\hardrange{2 -- 10 keV}
\def\softrange{0.5 -- 2 keV}
\def\hardfluxrange{$S_{\rm 2 - 10 keV}$}

\def\lognlogs{$\log N - \log S$}
\def\spose#1{\hbox to 0pt{#1\hss}}
\def\simlt{\mathrel{\spose{\lower 3pt\hbox{$\mathchar"218$}}
     \raise 2.0pt\hbox{$\mathchar"13C$}}}
\def\simgt{\mathrel{\spose{\lower 3pt\hbox{$\mathchar"218$}}
     \raise 2.0pt\hbox{$\mathchar"13E$}}}

\begin{document} 

\title{The Serendipitous Extragalactic X-Ray Source Identification (SEXSI)
Program: I. Characteristics of the Hard X-Ray Sample} 

\author{Fiona A. Harrison\altaffilmark{1},
Megan E. Eckart\altaffilmark{1},
Peter H. Mao\altaffilmark{1},  
David J. Helfand\altaffilmark{2},
Daniel Stern\altaffilmark{3}}

\begin{abstract} 
The Serendipitous Extragalactic X-ray Source Identification (SEXSI)
Program is designed to extend greatly the sample of identified
extragalactic hard X-ray (\hardrange) sources at intermediate fluxes
($\sim 10^{-13} - 10^{-15} \fluxu$).  SEXSI, which studies sources
selected from more than 2 deg$^2$, provides an essential complement to the
{\em Chandra} Deep Fields, which reach depths of $5 \times 10^{-16} \fluxu$
(\hardrange) but over a total area of $< 0.2$~deg$^2$.  In this paper we
describe the characteristics of the survey and our X-ray data analysis
methodology. We present the cumulative flux distribution for the
X-ray sample of 1034 hard sources, and discuss the distribution of spectral
hardness ratios. Our \lognlogs~in this intermediate flux range
connects to those found in the deep fields, and by combining the
data sets, we constrain the hard X-ray population over the flux 
range where the differential number counts change
slope, and from which the bulk of the 2 -- 10 keV X-ray background
arises.  We further investigate the \lognlogs~distribution separately for
soft and hard sources in our sample, finding that while
a clear change in slope is seen for the softer sample, the hardest sources
are well-described by a single power-law down to the faintest
fluxes, consistent with the notion that they lie at lower average redshift.
\end{abstract}
 
\altaffiltext{1}{Space Radiation Laboratory, 220-47, California Institute of 
Technology, Pasadena, CA 91125.} 
\altaffiltext{2}{Columbia University, Department of Astronomy, 550 West 120th Street, New York, NY 10027.} 
\altaffiltext{3}{Jet Propulsion Laboratory, California Institute of Technology, 
Mail Stop 169-506, Pasadena, CA 91109}

\section{Introduction}

A primary scientific motivation for developing the {\em Chandra X-ray
Observatory} was to perform surveys of the extragalactic sky up to
10~keV.  The combination of {\em Chandra's} superb angular resolving
power and high-energy response is enabling the detection and optical
identification of hard X-ray source populations at much fainter fluxes than
previously possible \citep{wov+96}.  Exposure times of 1~Ms in each
of two deep fields, the {\em Chandra} Deep Field-North \citep[CDF-N;][]{bah+01}
 and -South \citep[CDF-S;][]{gzw+02} reach depths of
$5 \times 10^{-16} \fluxu$ (\hardrange), and have
resolved most of the X-ray background up to 7~keV.  Optical spectroscopic
followup of a sample of the Deep-Field sources has revealed a diverse
counterpart population \citep{rtg+02,bcb+02}.  Attention is now
concentrated on understanding the physical nature of the counterparts,
as well as their evolution over cosmic time.

Wider field-of-view surveys provide an essential complement to the
Deep Fields -- which in total cover $<0.2$~deg$^2$ -- particularly for
this latter objective.  Large-area coverage is essential for providing
statistically significant source samples at intermediate to bright
fluxes (\hardfluxrange~$\sim 10^{-13} - 10^{-15} \fluxu$).  At
the bright end of this flux range there are only $\sim 20$ sources 
deg$^{-2}$, so that several square degrees must be covered to
obtain significant samples.  Spectroscopic identification of a
large fraction of these is necessary to sample broad redshift
and luminosity ranges, and to determine space densities of seemingly
rare populations such as high-redshift QSO II's, which appear about once
per 100~ksec {\em Chandra} field \citep{smc+02}.

Wide-field hard X-ray surveys undertaken with instruments prior to {\em Chandra}
and {\em XMM-Newton} made preliminary investigations of the bright end of
the hard source populations, although the positional accuracy achievable with
these experiments was insufficient to securely identify a large number of
counterparts.  The {\em BeppoSAX} HELLAS survey \citep{lfv+02} identified 61
sources either spectroscopically or from existing catalogs in 62 deg$^2$ to
a flux limit of \hardfluxrange~$= 5.0 \times 10^{-14} \fluxu$. The {\it
ASCA} Large Sky Survey identified 31 extragalactic sources in 20 deg$^2$
\citep{aoy+00}, with the recent addition of 85 more spectroscopically
identified sources from the {\it ASCA} Medium Sensitivity Survey \citep{auo+02}
to a flux threshold of \hardfluxrange~$= 1 \times 10^{-13} \fluxu$. 

{\em Chandra} and {\em XMM} offer the potential to expand 
significantly this initial
work to hundreds of sources detected over many square degrees.  A
number of programs are now underway to identify serendipitous sources
in extragalactic pointings.  Besides the {\em Chandra} Deep Fields, 
several other individual fields have
been studied, and optical spectroscopic followup completed and reported.  
These include the Lynx cluster field \citep{sts+02},
the field surrounding Abell 370 \citep{bcb+01}, and the Hawaii Survey Field
\citep{bcm+01}, each of which covers $\sim 0.08$~deg$^2$.  Ambitious
efforts to extend coverage to several deg$^2$ include the HELLAS2XMM
survey \citep{bmc+02}, and ChaMP \citep{h+02}, although followup results
from these surveys have not yet been published.

We present here the Serendipitous Extragalactic X-ray Source
Identification (SEXSI) program, a new hard X-ray survey designed to
fill the gap between wide-area, shallow surveys and the {\em Chandra}
Deep Fields. The survey has accumulated data from twenty-seven {\em
Chandra} fields selected from GTO and GO observations, covering more
than 2~deg$^2$.  We have cataloged more than 1000 sources in the
\hardrange~band, have completed deep optical imaging over most of the
survey area, and have obtained spectroscopic data on $\sim$~350 objects.

\begin{figure}
\epsscale{1} 
\plotone{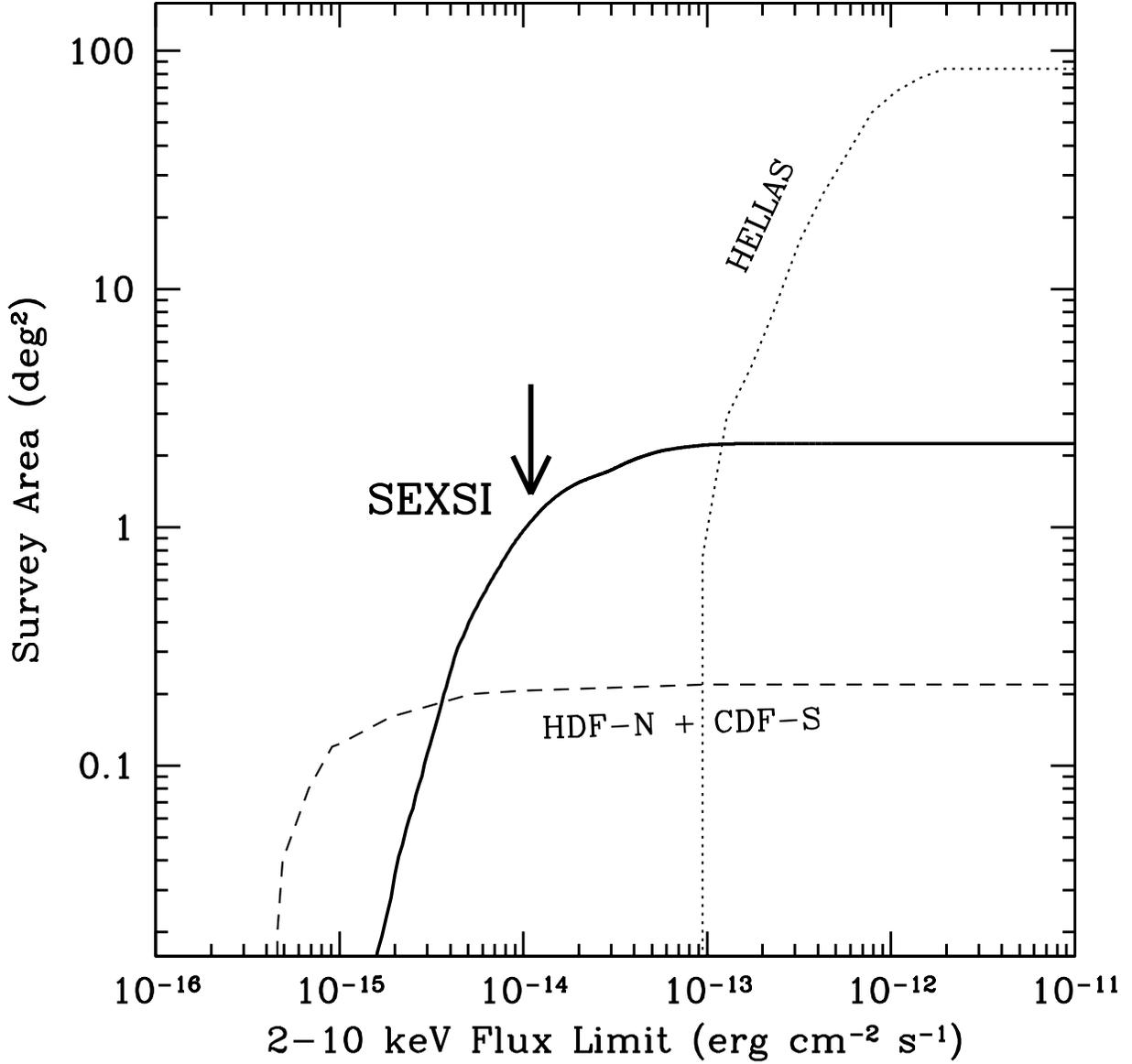} 
\caption{Area of sky surveyed as a function of hard (\hardrange) X-ray flux for
several hard X-ray surveys.  SEXSI (solid line) samples the parameter
space between the extremely deep, small area {\it Chandra} Ms
surveys in the Hubble Deep Field-North (HDF-N; Brandt et al. 2001) and
the Chandra Deep Field-South (CDF-S; Giacconi et al. 2002), and the
shallow, wide-area {\it BeppoSAX} High Energy Large Area Survey
(HELLAS; Fiore et al. 2001 -- dotted line).  Deep field coverage
(dashed line) corresponds to the 1~Ms depths, combining both fields.
HELLAS hard X-ray fluxes have been extrapolated from their published
5 -- 10 keV depths to \hardrange~ by multiplying by a factor of 1.96,
appropriate for the average $\alpha_E = 0.6$ they find in their
survey. The large arrow indicates the break in the \lognlogs~
plot (Figure~\ref{fig:difflnnlns}), corresponding to the flux level which dominates the
hard X-ray source counts.  The SEXSI survey is better-suited to exploring
this flux level than either the ultra-deep Ms {\it Chandra} surveys or the
shallow {\it BeppoSAX} HELLAS survey. \label{fig:areadepth}} 
 
\end{figure}

Table~\ref{tbl:surveycomp} summarizes the published \hardrange~X-ray 
surveys and demonstrates the
contribution of the SEXSI program. Tabulated flux values have been corrected
to the energy band \hardrange~and to the spectral assumptions adopted here
(see \S\ref{sec:reduction}), as detailed in the footnotes to the table. 
In the flux range 
$10^{-11}$ to $10^{-15} \fluxu$, the seven existing surveys have
discovered a total of 789 sources, several hundred fewer than the total 
presented here. In the flux range $10^{-13}$ to $3 \times 10^{-15} \fluxu$
which lies between the {\em ASCA} and {\em BeppoSAX} sensitivity
limits and the {\em Chandra} Deep Survey capability -- where the
\lognlogs~relation changes slope and from which the bulk of the
\hardrange~X-ray background arises
\citep{cgb+02} -- we more than triple the number of known sources. 
Figure~\ref{fig:areadepth} illustrates our areal coverage in
comparison with that of previous work, emphasizing how SEXSI 
complements previous surveys.

In this paper, we present the survey methodology and X-ray data
analysis techniques adopted, the X-ray source catalog, and the general
characteristics of the X-ray source sample. In a companion paper
(\citealt{els+03}; Paper II), we provide a summary of our optical
followup work, including a catalog of $R$ magnitudes (and upper limits
thereto) for over 1000 serendipitous X-rays sources as well as
redshifts and spectral classifications for $\sim 350$ of these
objects, and discuss the luminosity distribution, redshift
distribution, and composition of the sample.  Future papers will
address detailed analyses of the different source populations as well
as field-to-field variations.

\section{Selection of {\em Chandra} fields}

We selected fields with high Galactic latitude ($|b| > 20^{\circ}$)
and with declinations accessible to the optical facilities available
to us ($\delta > -20^{\circ}$).  We use observations taken with the
Advanced Camera for Imaging Spectroscopy (ACIS I- and S-modes; \citealt{bpb+98})
only (for sensitivity in the hard band).  All the fields presented in
this paper have data which are currently in the {\em Chandra} public
archive, although in many cases we made arrangements with the target
PI for advanced access in order to begin spectroscopic observations
prior to public release of the data.  Table~\ref{tbl:fields} lists the
27 survey fields by target, and includes the target type and redshift if
known, the coordinates of the field center, the Galactic neutral
hydrogen column density in this direction \citep{dl+90}, the X-ray
exposure time, and the ACIS chips reduced and included in this
work. The observations included in our survey represent a total of
1.65 Msec of on-source integration time and include data from 134
$8^{\prime} \times 8^{\prime}$ ACIS chips.

\begin{figure}
\epsscale{1} 
\plotone{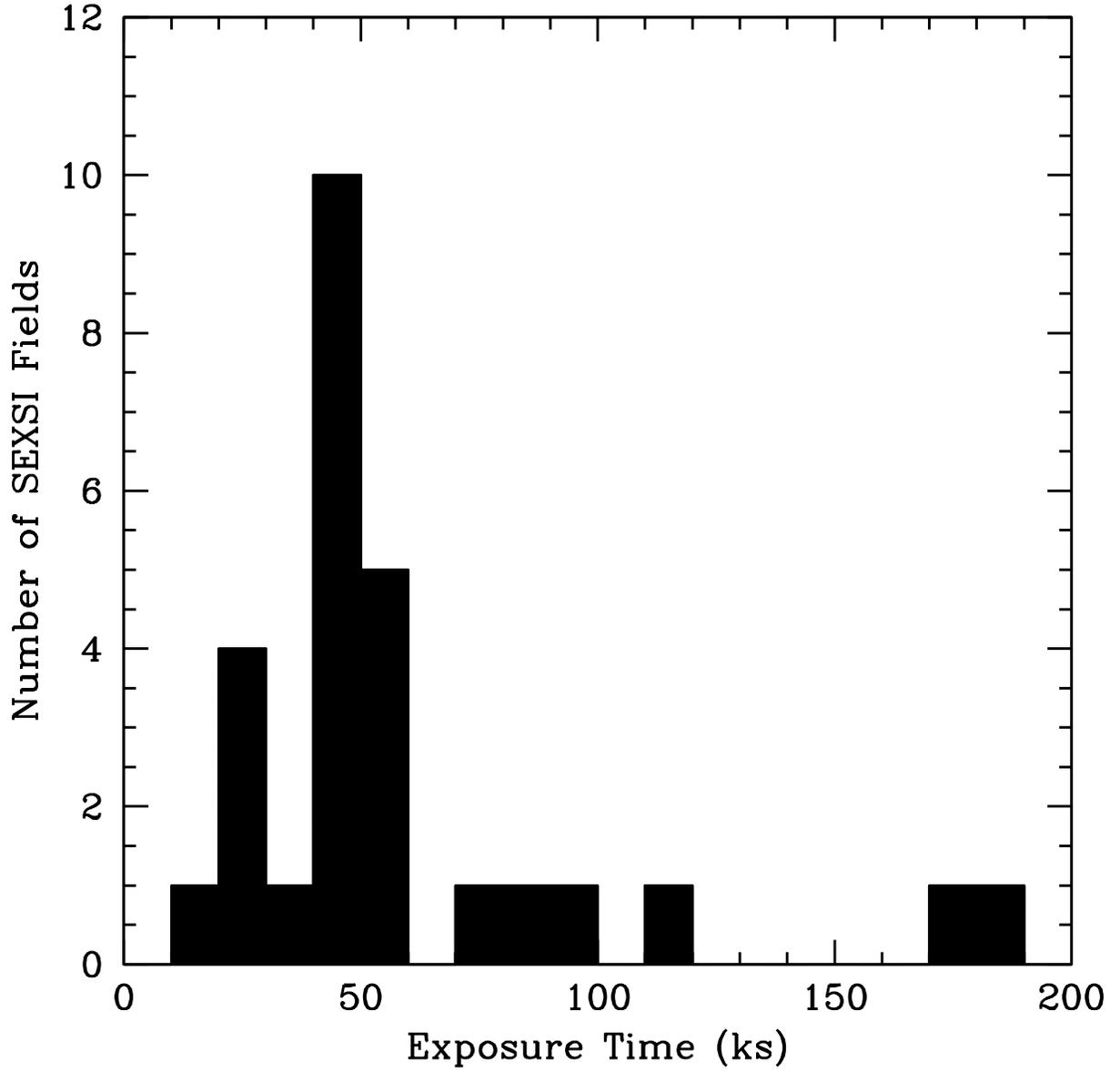} 
\caption{Histogram of exposure times for the 27 {\it Chandra} fields listed in Table~\ref{tbl:fields}. \label{fig:exptimes}} 
 
\end{figure}

Net exposure times range from 18 ks to 186 ks; a histogram of exposure
times is given in Figure~\ref{fig:exptimes}. Targets include a
Galactic planetary nebula, various types of AGN, transient afterglow
followup observations, NGC galaxies, and clusters of galaxies,
particularly those at relatively high redshifts. For the cases in
which the target is an extended X-ray source, we have taken care to
exclude those sources potentially associated with the target from our
\lognlogs~analysis (see \S\ref{sec:srcdeletions}).

\section{Data reduction and analysis}

\subsection{Basic X-ray reduction}
\label{sec:reduction}

The X-ray data reduction includes filtering raw event data to reject
contaminating particle events, binning the event data into images with
specific energy ranges, searching the images for sources, and extracting
source fluxes.  

For the initial processing steps, through source identification, we
use standard tools supplied by the {\em Chandra X-ray Observatory}
Science Center (CXC).  We employ {\em ASCA} event grades 0, 2, 3, 4,
6, and we eliminate flickering pixels and events at the chip node
boundaries.  For each chip we bin events into soft (0.3 -- 2.1 keV)
and hard (2.1 -- 7 keV) band images. The 2.1~keV energy boundary is
chosen to coincide with the abrupt mirror reflectance change caused by
the Ir M-shell edge, and the upper and lower limits optimize signal-to-noise 
in the images. We use {\tt wavdetect} for initial source identification.
In a subsequent step we test the significance of each source individually 
and eliminate sources with a nominal chance occurrence probability 
greater than $10^{-6}$.

For the remainder of the processing, we use primarily our own routines
to filter the {\tt wavdetect} source list to reject spurious
detections, to extract source fluxes, and to correct {\tt wavdetect}
positions when required. In some cases, particularly at large off-axis
angles where the point spread function (PSF) is relatively broad, the
{\tt wavdetect} positions become unreliable, with some positions
differing significantly from the centroid of the photon distribution.
The differences are not uniformly distributed, and most are within the
expected statistical tolerance. However, in typically one or two cases
per field the {\tt wavdetect} position will differ unacceptably, a
discrepancy that has also been noted by others \citep{bah+01}.  We use
the {\tt wavdetect} positions, the standard used by most other
authors, unless the PSF-normalized radial shift $\Delta r^2/{\rm PSF}
> 0.8$ arcsec, in which case we use the centroid position.

After correcting source positions, we extract photons from the image
to determine source fluxes. The PSF width is a strong function of
off-axis angle. To determine extraction radii, we use the encircled
energy fractions tabulated by the CXC at eight off-axis angles
and at five different energies. We use the 1.5 keV and 4.5 keV values
for the soft and hard bands, respectively, and interpolate linearly
between tabulated values in off-axis angle.  For the extraction radius
we use an encircled energy fraction ranging from 80 -- 90\%, depending
on the band and off-axis angle (see Table~\ref{tbl:psfoaa}). This optimizes the
signal-to-noise ratio, since the optimal fraction depends on signal-to-background 
ratio. To determine the background level for subtraction,
we identify a number of circular, source-free regions on each chip,
and for each source, use the closest region to determine the
background.  We define a sufficient number of regions distributed over
the chip to ensure that systematic background variations are small
compared to statistical uncertainties.

For each {\tt wavdetect} source, we use the background level in 
the extraction aperture to calculate a lower limit to the number 
of total counts for which the probability\footnote{The probability 
is calculated using the Poisson distribution for low-count ($< 33$) 
sources and the Gaussian limit for high-count ($> 33$) sources.} 
that the detection is a random fluctuation 
is less than $10^{-6}$. If the total extracted counts fall below 
this limit, we deem the candidate {\tt wavdetect} source to have 
failed our significance criterion, and remove the source from the 
catalog. In on-axis chips, there are about $5 \times 10^{5}$ detection 
cells, so we expect $\sim 0.5$ false detections per chip.  Off-axis 
chips have 4 -- 8 times fewer detection cells, as we bin them before 
searching.  Thus, on average we expect $\leq 1$ false detection per 
field, depending on the number and configuration of chips read out.

To convert extracted source counts to flux, we use standard {\em CXO}
software to compute energy-weighted exposure maps using a power-law spectral 
model with photon spectral index $\Gamma = 1.5$. Using these, we convert
soft band counts to a 0.5 -- 2 keV flux, and hard band counts to a 2
-- 10 keV flux again adopting $\Gamma = 1.5$, and apply an aperture 
correction to account for the varying encircled energy fraction used 
in source extraction (see Table~\ref{tbl:psfoaa}). We use the Galactic column
density for each field listed in Table~\ref{tbl:fields} to calculate
source fluxes arriving at the Galaxy in the hard and soft bands.  
For an on-axis source, the conversion factor in the hard band is
\hardfluxrange~$\sim 3.0 \times 10^{-11} \fluxu$ ct$^{-1}$, although this value
varies by 5-10\% from field to field owing to the differences in
location of the aim
point relative to node and chip boundaries.  We note that $\Gamma =
1.5$ represents a softer spectrum than the $\Gamma = 1.2 - 1.4$
typically used for counts to flux conversion for the deep fields; our choice was
motivated by the brighter average flux of our sample.

\subsection{Source deletions}
\label{sec:srcdeletions}

In order to calculate the \lognlogs~relation and to characterize the
serendipitous source populations in an unbiased manner, we remove
sources associated with the observation targets. In the case of point
source targets such as AGN, transient afterglows, and the planetary
nebula, this excision is trivial: the target object is simply excluded
from the catalog.  For the nearby galaxies in which the target covers
a significant area of the field, we have excised all sources within an
elliptical region defined where, in our optical image of the galaxy,
the galactic light is $>$ 11$\sigma$ above the average background
level.  This led to the removal of 68 sources from the
catalog. Finally, in the case of galaxy clusters, Chandra's high
angular resolution allows one to easily `see through' the diffuse
emission from the hot intracluster gas to the universe beyond, and it
is not necessary to exclude all discrete sources for optical followup
studies. Some such sources are, however, associated with the target
cluster, and should not be included in our \lognlogs~analysis.
Thus, apart from a few sources detected in the hard band which represent the 
diffuse cluster emission (and have thus been removed from the catalog),
we have included all discrete sources detected in the cluster fields,
but have flagged all those within $\sim 1$ Mpc of the cluster centroid
as potentially associated with the target.  We exclude these flagged sources 
(and the associated effective area) from the \lognlogs~analysis (a total of
190 sources).  Only a small fraction of these are actually
optical spectroscopically-identified cluster members.

\subsection{Hardness ratio calculation}
\label{sec:hrcalc}

We define the hardness ratios as $(H-S)/(H+S)$, where H and S are the
corrected counts in the \hardrange~and \softrange~bands respectively.
We extract the counts from our hard- (2.1--7 keV) and soft- (0.3--2.1 keV) 
band images using the centroids obtained by running {\tt wavdetect} on the 
images separately (and subsequently correct the rates to the standard bands).
We do this, rather than extracting counts from the
soft-band image using the hard-band positions in order to minimize
bias, as described below.

\begin{figure}
\epsscale{1} 
\plotone{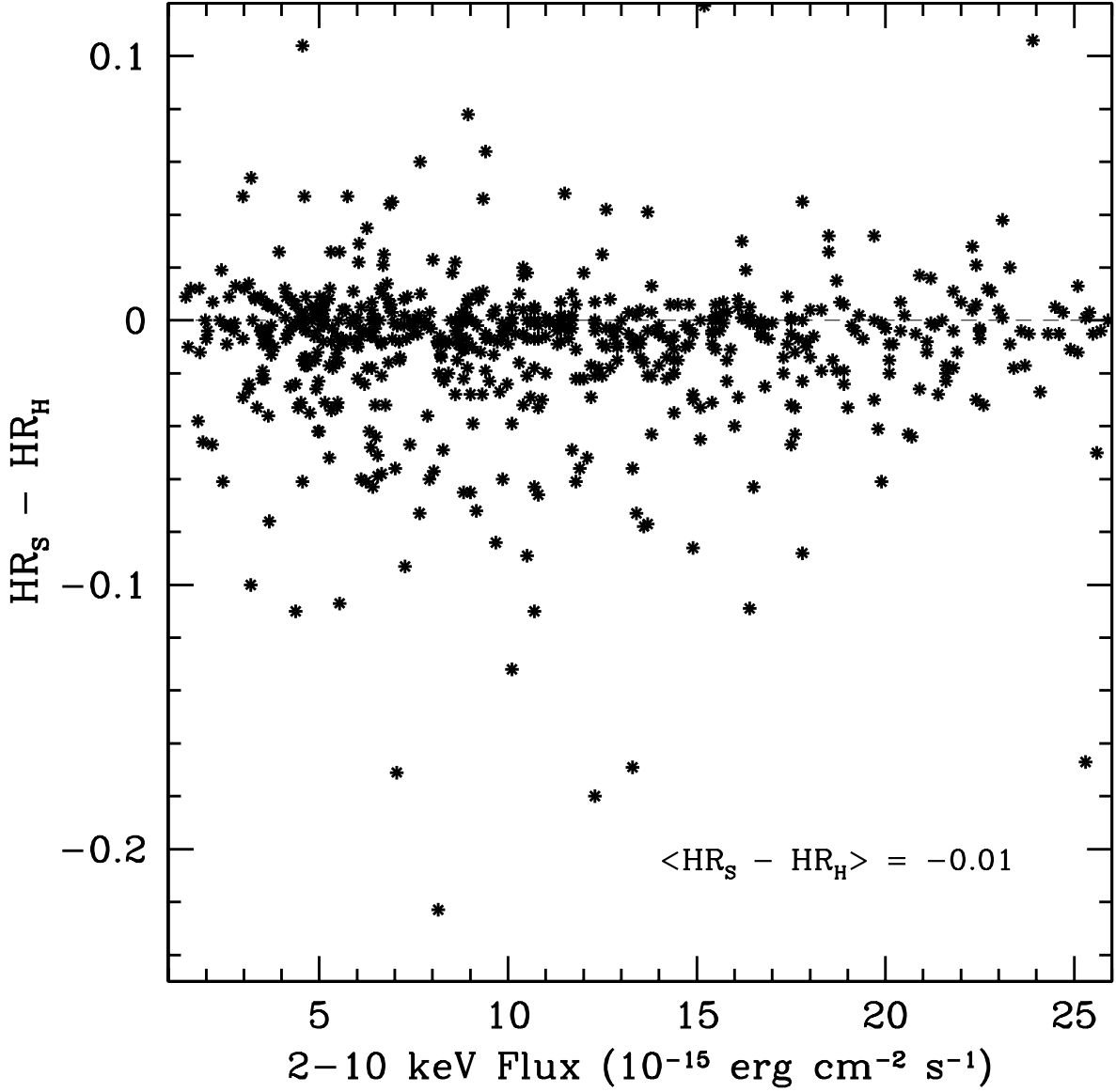} 
\caption{Comparison of hardness ratio $HR \equiv (H-S)/(H+S)$ as a function of
hard X-ray flux ($H$; \hardrange), with soft-band flux ($S$;
\softrange) measured in two ways. We exclude hard-band only sources;
 they are plotted separately in Figure~\ref{fig:hrcomp.eq0}.  
Our primary methodology (see \S\ref{sec:hrcalc})
 is to produce soft-band catalogs directly from the soft-band
images using {\tt wavdetect}.  The matched soft-band and hard-band
catalogs were then used to produce {\em soft-band registered} hardness
ratios $HR_S$; these are the hardness ratios presented as $HR$ in
Table~\ref{tbl:catalog} and Figures~\ref{fig:hardness} and~\ref{fig:hardhist}.
Alternatively, hardness ratios were derived
by extracting the soft-band flux using the aperture defined by the
hard-band detection; the resultant {\em hard-band registered} hardness
ratios are indicated here as $HR_H$.  This figure shows that the
difference between the techniques is typically less than a 0.1 in $HR$
for any given source, with only a very slight systematic for $HR_H$
producing more positive hardness ratios.\label{fig:hrcomp.n0}} 
 \end{figure}

\begin{figure}
\epsscale{1} 
\plotone{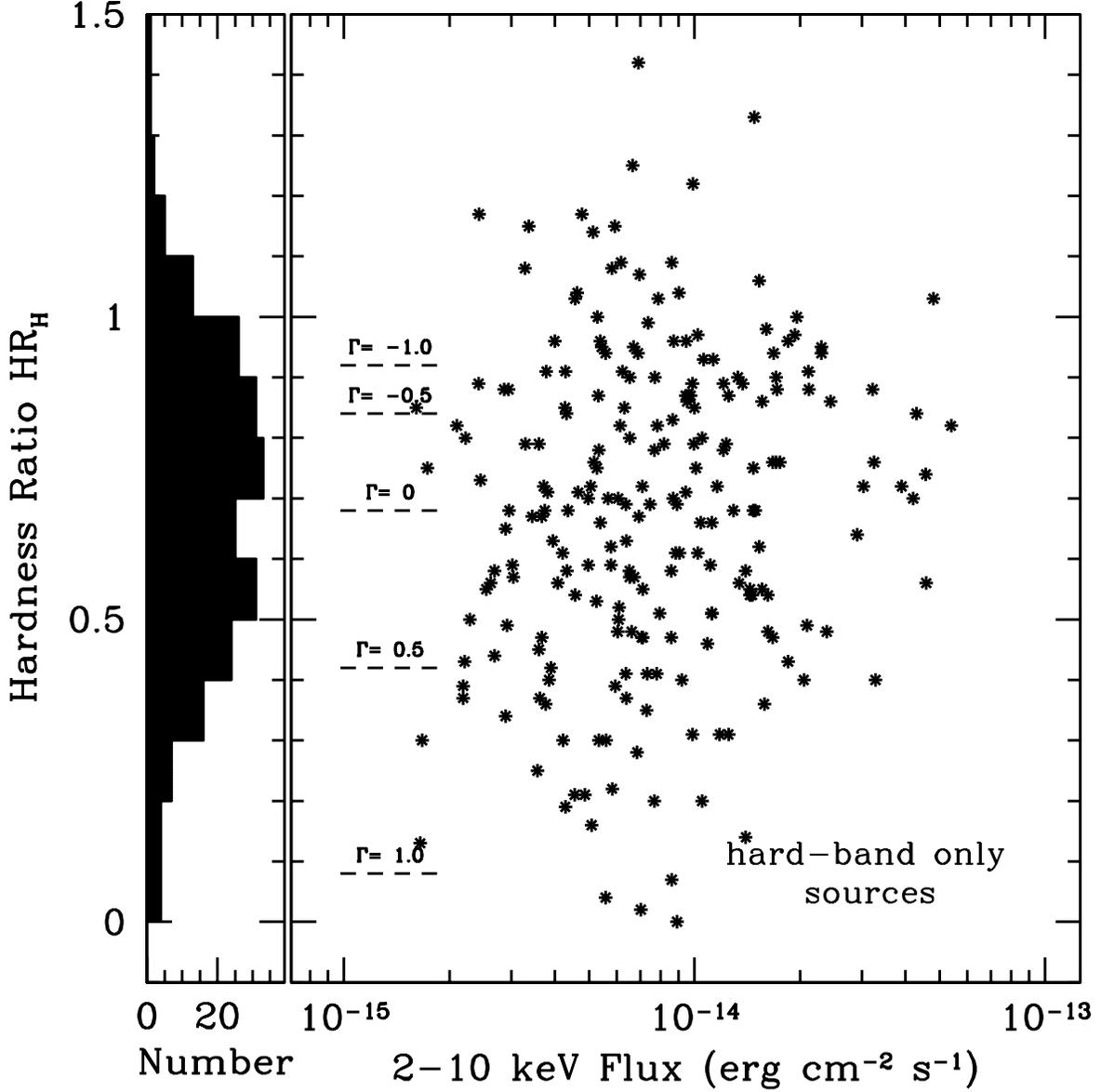} 
\caption{Hardness ratio $HR_H \equiv (H-S)/(H+S)$ as a function of hard X-ray
flux for sources lacking a counterpart in the soft X-ray catalog.  The 
left-hand panel collapses the scatter plot, showing the histogram of 
hard-band registered hardness ratios. The dashed horizontal lines indicate
the photon index $\Gamma$ which different hardness ratios correspond to.
In Figure~\ref{fig:hardness} these sources are all plotted along the horizontal line
corresponding to $HR = 1$.  Here we extract soft-band counts using the
aperture defined by the hard-band detection; when the extracted
soft-band counts are negative, the resultant {\em hard-band registered}
hardness ratio $HR_H$ is greater than unity. Visual inspection of the 
soft-band images of all sources with \hardfluxrange~$\geq 2.5 \times 10^{-14} 
\fluxu$ verifies the lack of soft-band detections. \label{fig:hrcomp.eq0} } 
\end{figure}

In a small number of cases, {\tt wavdetect} failed to find a soft
source both clearly present in the image and with a hard source
counterpart (typically as a result of a second source very nearby). In
order to correct these discrepancies, and to test for any systematic
differences in soft and hard source positions, we also derived a soft
flux for each source using the hard source centroid. We calculated
hardness ratios using both sets of soft counts (those
derived by {\tt wavdetect} and those extracted using hard-source
positions). Figures~\ref{fig:hrcomp.n0} and~\ref{fig:hrcomp.eq0} show a
comparison of the two techniques.

Using the optimal centroid position for a fixed aperture to extract
source counts systematically overestimates source fluxes since the
centroid selected will be influenced by positive background
fluctuations to maximize the number of counts included -- a form of
Eddington bias; see \citet{cgb+02}. Thus, calculating hardness
ratios by using soft counts extracted from a region with the hard
source centroid will produce a systematic bias toward greater hardness
ratios. This is illustrated in Figure~\ref{fig:hrcomp.n0}, where we
show the difference in the two methods for calculating the hardness
ratio as a function of soft source flux; while only 9 sources have a
difference of $> +0.05$, 48 sources have a difference of $<
-0.05$. The mean bias is $-0.01$. To avoid this, we use the hardness
ratios derived from the independent soft and hard source catalogs.

\subsection{Calculation of the effective area function}

In order to construct the hard X-ray source \lognlogs~curve, we must determine
the effective area of our survey as a function of source flux (shown in
Figure~\ref{fig:areadepth}). We do this by using the same algorithms we employ
for the actual source extraction and flux conversion, calculated on a fine grid
which samples the entire field of view.  Our detailed calculation assures
that, independent of the methodology used for background subtraction
and source significance testing, our calculation of the effective area
will be accurate.  Also, since we employ significant off-axis area in
the survey, calculating the response with fine sampling across the
field of view is required, given the rapid PSF changes with 
off-axis angle and telescope vignetting.

We divide the images from each chip, with the detected sources
`blanked out', into a fine grid sampled at a pitch of 8 pixels.  At
each location, we repeat the steps associated with source detection:
we determine the aperture from the off-axis angle, background from the
closest circular background region, and effective area from the
spectrally-weighted exposure map at that location.  Using these, we
determine the minimum detectable flux at that location corresponding
to a spurious detection probability of $10^{-6}$.  We step across the
grid in this manner, so that we determine accurately the sky area as a
function of minimum detectable flux even for chips where the response
changes significantly over the image. This procedure results in an effective
area function which optimally matches the source detection procedure
and supports the construction of a \lognlogs~curve free of any biases which
might be introduced by the approximate techniques adopted by some
other surveys.

\section{The source catalog}
\label{sec:catalog}

In Table~\ref{tbl:catalog} we present the SEXSI source catalog of 1034
hard-band discrete serendipitous X-ray sources detected as described
above.  Sources are designated CXOSEXSI (our IAU-registered name)
followed by standard truncated source coordinates. The source
positions (equinox J2000.0) are those derived from the hard-band X-ray
images; in Paper II we use optical astrometry to derive mean offsets
for each X-ray image and provide improved positions (although offsets
are typically less than $1^{\prime\prime}$). We include only sources
detected with a chance coincidence probability of $<10^{-6}$ in the
hard band. The angular distance of the source position from the
telescope axis is given in column 4.  Columns 5 and 6 list the
background-subtracted counts for each source within the specified
aperture derived from the 2.1 -- 7 keV image, followed by the
estimated background counts in that same aperture. Column 7 gives an
estimate of the signal-to-noise ratio (SNR) of the detection. The SNR
is calculated using the approximate formula for the Poisson
distribution with small numbers of counts given in \citet{g+86}:

\begin{equation}
{\rm SNR} = \frac{{\rm source~counts}}{1 + \sqrt{0.75 + {\rm source~counts + background~counts}}}~;
\label{eqn:SNR}
\end{equation}

\noindent
for high-count sources Equation~\ref{eqn:SNR} converges to the Gaussian
limit.  Owing to the relatively large background regions we have
employed, the background error is negligible in the SNR
calculation. It should be emphasized that these values are {\it not} a
measure of source significance (which is $P<10^{-6}$ in all
cases) but is a measure of the uncertainty in the source flux
estimates. Column 8 shows the unabsorbed hard band flux (in units of $10^{-15} \fluxu$), 
corrected for source counts falling outside the aperture and translated to the
standard \hardrange~band assuming a power law photon spectral index of $\Gamma
=1.5$ and a Galactic absorbing column density appropriate to the field
(see Table~\ref{tbl:fields}). 
Columns 9 -- 12 provide the analogous values for the soft band. We derived
soft-band fluxes by employing the same procedures on the {\tt
wavdetect} output from the 0.3 -- 2.1 keV images, and then matching
sources in the two bands. The soft fluxes are presented in the \softrange.
There are a large number of soft sources
which lack a statistically significant hard counterpart; however, as
we are interested in the \hardrange~source populations, these sources
are not included in the Table or considered further here. We do include
a catalog of the soft band only sources in the Appendix.

\subsection{Comparison of methodology with previous work}
\label{sec:compare}

As \citet{cgb+02} have recently discussed, the details of source
detection and flux extraction can have non-trivial effects on the
final source catalog derived from an X-ray image, as well as on
conclusions drawn from the \lognlogs~relation. As a test of our
methodology, we have compared our results on one of the deeper fields
in our sample, CL0848+4454, with the analysis published by 
\citeauthor{sts+02} (2002b; the SPICES survey) which uses the 
same technique as \citet{gzw+02} apply to the CDF-S. The source detection
algorithm, the flux estimation method, and the effective area
calculations all differ from ours, so a comparison is instructive.

\citet{sts+02} use the SExtractor source detection algorithm
\citep{ba+96} applied to a version of the 0.5 -- 7 keV image with a
smoothed background and a signal-to-noise cut of 2.1. They measure
source fluxes using a source aperture of $R_S= 2.4~\times$ the PSF
full-width at half-maximum, with the background derived from an
annulus $R_S+2^{\prime\prime}$ to $R_S+12^{\prime\prime}$.  Their 
simulations predict five false sources using this procedure.  In contrast, 
we employ the {\tt wavdetect} algorithm to generate a list of source candidates from
the 2.1 -- 7 keV image, use hand-selected, source-free background regions
with larger average areas to minimize statistical uncertainties, and require 
each source to have a probability of chance occurrence $<10^{-6}$, yielding 
$<1$ false source in this field. As noted above, we also calculate a fine-scale
effective area function using exactly the same significance criterion
for each PSF area on the image. 

\begin{figure}
\epsscale{1} 
\plotone{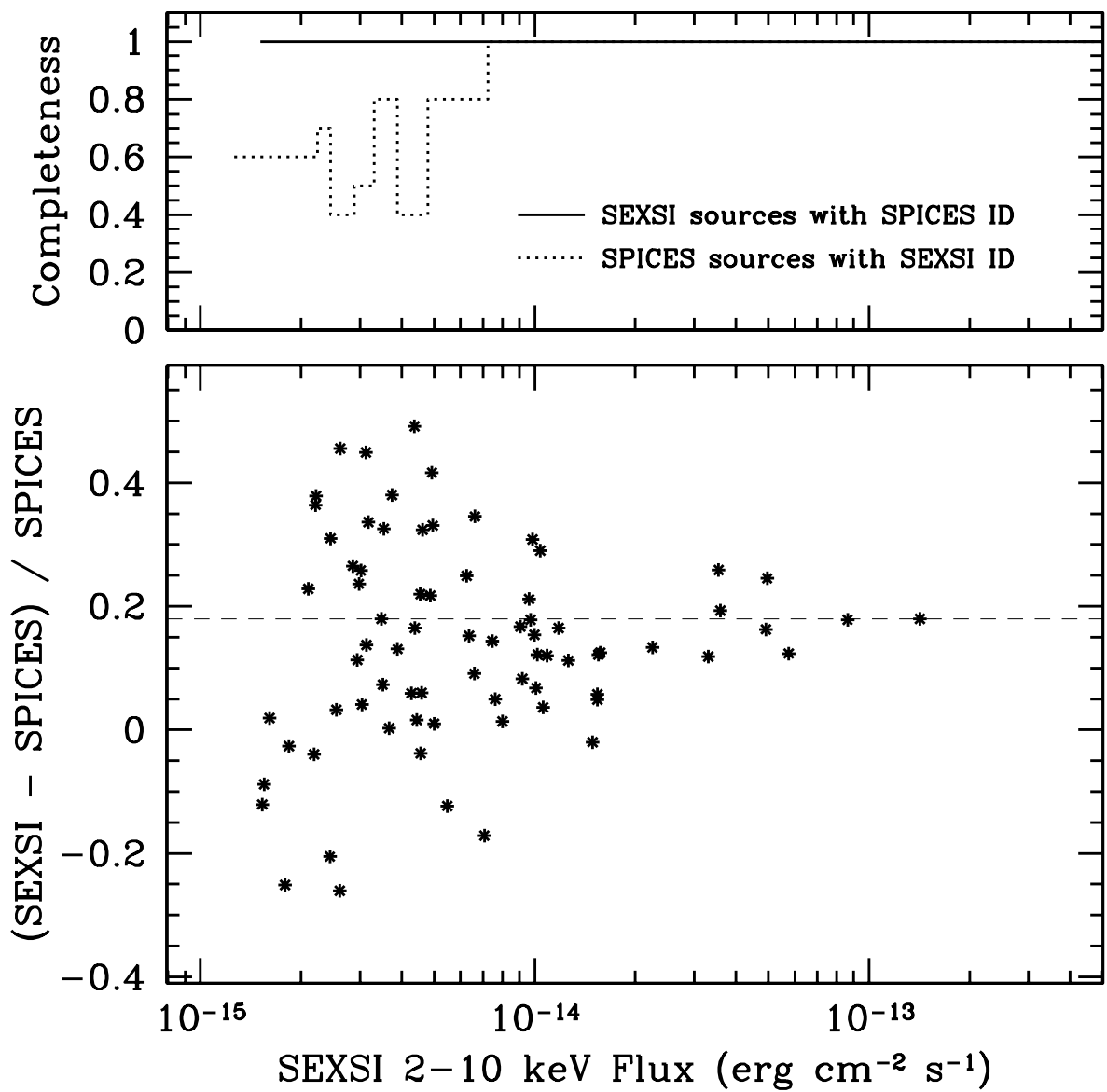} 
\caption{Comparison of the SEXSI and SPICES~\citep{sts+02} catalogs of
hard-band \hardrange~sources from the 185~ks {\it Chandra}/ACIS
observation of the CL0848+4454 field in Lynx.  The top panel shows the
fraction of sources in each catalog which are identified in the other
catalog, plotted with 10 sources per histogram bin.  At bright fluxes,
the source catalogs are identical.  At lower fluxes, SPICES includes 33
sources that SEXSI does not, likely due to the different detection algorithms and
detection passbands.  The bottom panel compares photometry for the 78
sources which appear in both catalogs:  SEXSI hard-band fluxes are
systematically $\approx 18\%$ higher (dashed horizontal line).\label{fig:lynx} } 
\end{figure}

Figure~\ref{fig:lynx} summarizes the result of comparing the two
source catalogs. Apart from three bright sources in the off-axis S-6
chip which was not analyzed by Stern et al., our catalogs are
identical down to a flux threshold of $10^{-14} \fluxu$. At fainter
fluxes, there is a large number of SPICES sources -- 33 hard band
detections -- which fail to appear in our catalog (see
Figure~\ref{fig:lynx} -- upper panel). We examined the hard-band
images at each of these locations. In seven cases, our {\tt wavdetect}
algorithm indicated source candidates were present, but each failed
the $P<10^{-6}$ significance test in the hard band. In most of the
other cases, no source was apparent in the hard band, although quite a
number had soft-band counterparts. In some cases, fortuitous
background fluctuations in the annulus surrounding the source may have
accounted for the reported SPICES hard-band detection. In nearly a
third of the cases, a plausible optical identification has been found,
so it is clear that some of these sources are real X-ray
emitters. However, in no case did our algorithm suite miss a source
which would pass our specified threshold. Since our effective area
function is calculated in a manner fully consistent with our threshold
calculation, our \lognlogs~will be unaffected by the absence of
these faint, low-significance sources from our catalog.

The lower panel of Figure~\ref{fig:lynx} indicates a systematic offset
between the flux scale of the two catalogs of 15 -- 20\%, with our
SEXSI fluxes being systematically higher. Most of this effect is
explained by the count-to-flux conversion factors adopted in the two
studies (3.24 and $2.79 \times 10^{-11} \fluxu$ ct$^{-1}$ respectively
for SEXSI and SPICES), which in turn derives from the use of slightly
different spectral index assumptions and a different generation of
response function for the instrument. Individual fluxes for weaker
sources have discrepancies of up to 40\%, which can be accounted for by
different flux extraction and background subtraction algorithms
applied to low-count-rate sources.

In summary, the differences between the two analyses of this field,
while producing catalogs differing at the $\sim 20\%$ level in both
source existence and source flux, are well understood. In particular
we are confident that the self-consistent method we have adopted for
calculating the source detection threshold and the effective area
function will yield an unbiased estimate of the true \lognlogs~
relation for hard-band X-ray sources.

\section{The \hardrange~$\mathbf{\log N - \log S}$ relation}
\label{sec:lognlogs}

The CDF-N and -S have provided good measurements of the \hardrange~
\lognlogs~relationship at fluxes below $\sim 10^{-14} \fluxu$ .  In
comparison, the SEXSI sample includes 478 sources with fluxes between
$10^{-12}$ and $10^{-14} \fluxu $.  By combining our measurements with
the deep field results, we can constrain the \lognlogs~over a broad
range, which includes the break from Euclidean behavior. 

We use the CDF-S fluxes from \citet{gzw+02} along with the SEXSI
sample to construct the \lognlogs~between $10^{-12}$ and $10^{-15}
\fluxu$.  For the same reasons given by \citet{cgb+02}, we choose to
work with the differential curve: the differential measurement
provides statistically-independent bins, and comparison does not rely
on the bright-end normalization, which must be taken from other
instruments.  To calculate the SEXSI \lognlogs, we use the effective
area curve (Figure~\ref{fig:areadepth}) to correct for incompleteness
at the faint end of the sample.  We have not corrected for Eddington
bias which is, by comparison, a small effect.  We employ the CDF-S
fluxes with a correction (of about 5\%) to account for the different
spectral index assumption ($\Gamma= 1.375$ for CDF-S compared to
$\Gamma = 1.5$ for SEXSI).  To correct for incompleteness in the CDF-S
sample, we use the effective area curve provided to us by P. Tozzi.
We calculate the differential counts by binning $(N(S))$, the number
of sources with flux $S$, into flux ranges, $\Delta S_i$, then
computing the average effective area, $A_i$ for that range, and
forming the differential curve by
\begin{equation}
 n(S)_i = \sum_{S_{jmin}}^{S_{jmax}}(N(S_j))/(\Delta S_i A_i).
\end{equation}
We normalize to a unit flux of $10^{-14} \fluxu$.

\begin{figure}
\epsscale{1} 
\plotone{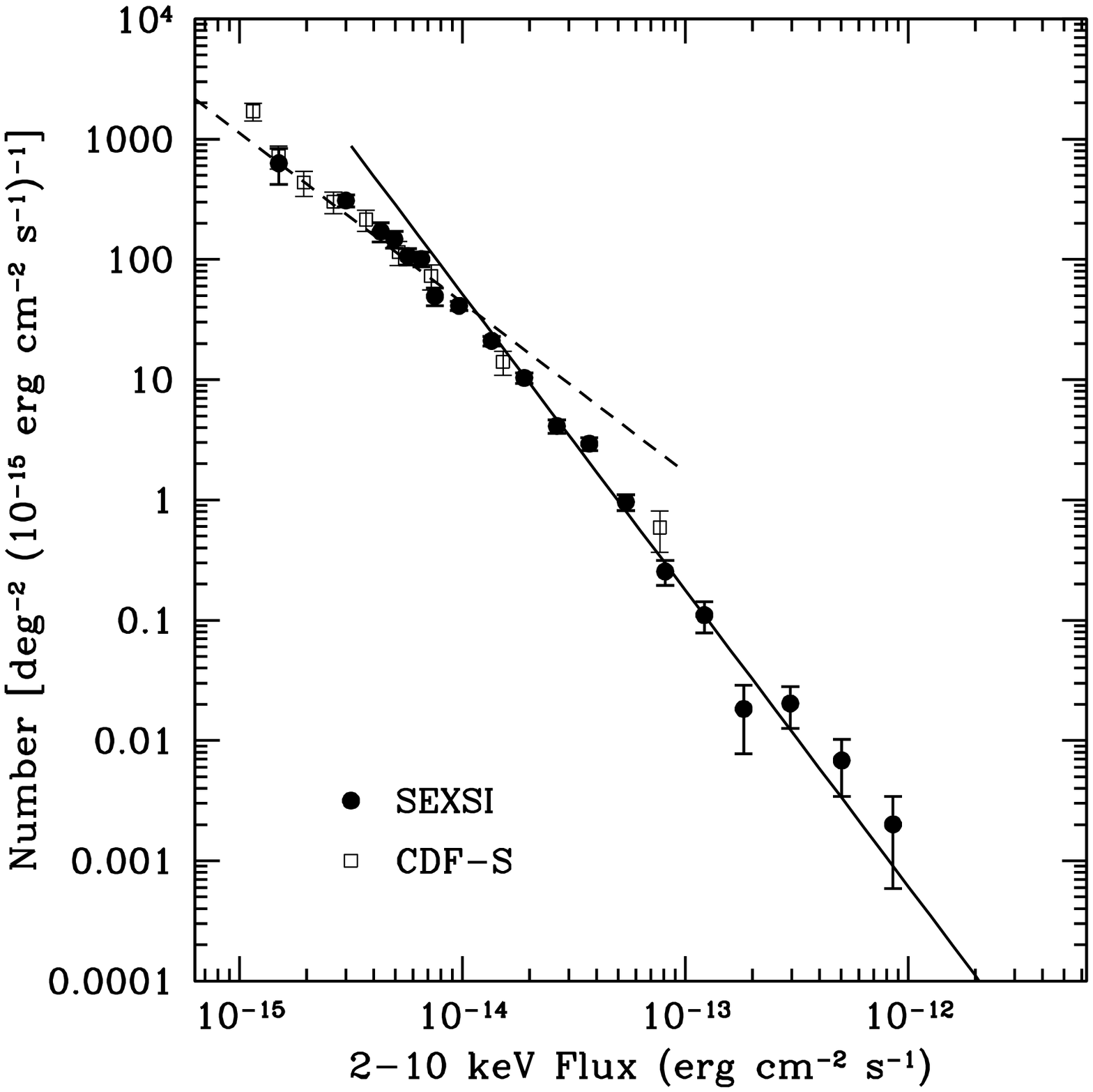} 
\caption{Differential \lognlogs~for the SEXSI survey (circles) and
Chandra Deep Field South (squares).  The data are plotted in units
of number deg$^{-2}$ per unit flux, where the flux has been divided
by $1 \times 10^{-15} \fluxu$.   The solid lines shows a linear
fit to the SEXSI data between $1 \times 10^{-12} \fluxu$
and $1 \times 10^{-14} \fluxu$, and the dashed line shows a
fit to the CDF-S data between $1 \times 10^{-14} \fluxu$
and $1 \times 10^{-15} \fluxu$.  The combined curve clearly
changes slope at $1 - 2 \times 10^{-14} \fluxu$.  Note that
the agreement of the normalization between the SEXSI and CDF-S data is good.
\label{fig:difflnnlns} } 
\end{figure}

Figure~\ref{fig:difflnnlns} shows the differential \lognlogs~curve
from the combined SEXSI and CDF-S catalogs,
where the indicated errors are 1$\sigma$.  The normalizations between
the two agree well in the region of overlap, especially considering
the different source extraction techniques and methodologies for
calculating the effective area function.  The combined data cannot be
fit with a single power law, but require a break in slope between $1 - 2
\times 10^{-14} \fluxu$.  We fit the SEXSI data with a single power
law at fluxes above $1.25 \times 10^{-14} \fluxu$, and the CDF-S data
to a separate power law below this.  The fits are shown as solid and
dashed lines in Figure~\ref{fig:difflnnlns}.  The two intersect at a
flux of $1.1 \times 10^{-14} \fluxu$.  We note that the exact position
of the intersection depends on where we divide the data, but for
reasonable choices yielding good fits, the break always lies in the
range $1 - 2 \times 10^{-14} \fluxu$ which contains the break point
first predicted on the basis of a fluctuation analysis of the {\it
Einstein} Deep Survey fields nearly two decades ago \citep{hh+87}.

The best-fit curves are parameterized by
\begin{equation}
n(S) = (46.8 \pm 2.1)(S/10^{-14})^{-2.46 \pm 0.08}
\end{equation}
for $S > 1.25 \times 10^{-14} \fluxu$, and
\begin{equation}
n(S) = (43.65 \pm_{2.0}^{2.1})(S/10^{-14})^{-1.41 \pm 0.17}
\end{equation}
below $1 \times 10^{-14} \fluxu$.  The quoted errors are $1\sigma$
formal errors on the fits.  The errors on the data points are
statistical errors only, and do not include an estimate of the
systematic uncertainties, such as biases on approximations in
correcting for incompleteness.  Based on the good agreement of
the overall normalization with other surveys (see below), the 
systematic errors do not exceed the statistical uncertainty.
The faint-end slope is dependent on where we divide the fit
ranges; cutting the data at $2.5 \times 10^{-14} \fluxu$ yields
an acceptable faint-end fit, but with a steeper slope of $-1.7$.

\begin{figure}
\epsscale{1} 
\plotone{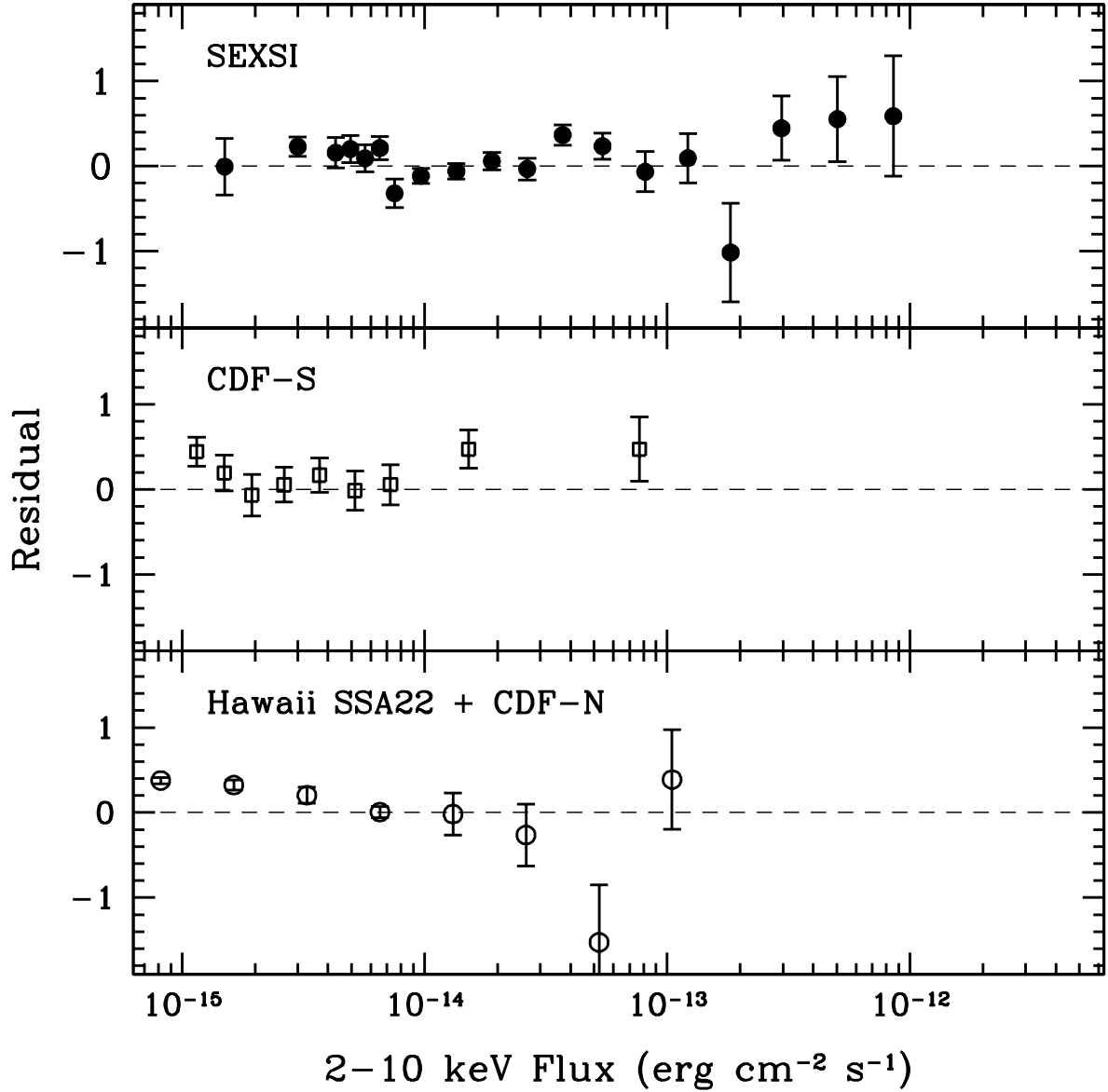} 
\caption{The residuals, defined as (data - fit)/data, for SEXSI,
CDF-S, and Hawaii SSA22 and CDF-N. The fit is our best fit to the
SEXSI data at the bright end, and the CDF-S data at the faint end. \label{fig:lnncomp}} 
\end{figure}

Figure~\ref{fig:lnncomp} shows the fractional residuals from the
best-fit curves for the SEXSI survey (top panel), the CDF-S (middle),
and combined Hawaii SSA22 and CDF-N sample \citep{cgb+02}.  For the
Hawaii/CDF-N data, we use the binned points (provided in digital form
by L. Cowie), corrected for the different spectral slope assumed for
counts to flux conversion.  At the faint end, the overall
normalizations agree reasonably well, with the CDF-N data
systematically slightly (1$\sigma$) above the mean fit to the SEXSI
and CDF-S data.  The faint-end slope of $-1.41 \pm 0.17$ found here is
marginally steeper than the best-fit values of $-1.63 \pm 0.054$ found
by \citet{cgb+02} and $-1.61 \pm 0.10$ found by \citet{rtg+02}.  
This difference is largely due to the somewhat
different normalization; in addition, as noted above, the placement of 
the power-law break and the binning affects the best-fit slope, so this discrepancy
is not significant. Above $2 \times 10^{-14} \fluxu$, the deep fields
contain only 2 -- 3 bins, and so the shape is much better constrained
by the SEXSI data.  Our best-fit slope at the bright end is $-2.46 \pm
0.08$, consistent both with a Euclidean source distribution, and with
the value of $-2.57 \pm 0.22$ found for the Hawaii+CDF-N data.

\section{X-ray properties of the sample}
\label{sec:xrayprops}

Most SEXSI sources have too few X-ray counts to warrant spectral
fitting, so we rely upon hardness ratios ($HR$) to characterize
the spectral slope. As discussed in
\S\ref{sec:hrcalc}, we calculate the hardness ratio for each source,
listed in column 13 of Table~\ref{tbl:catalog}, using positions
determined by independently searching the hard and soft images.  We
assign hard-band sources that have no soft-band {\tt wavdetect}
counterpart at our significance level a $HR$ of 1.0. We have also
determined a hardness ratio derived by extracting flux from the
soft-band images at the position determined by searching the hard-band
images, which we designate by $HR_H$.  Note this does not require a
significant independent detection in the soft band, so that for many
sources with $HR = 1$, $HR_H < 1$ (see Figure~\ref{fig:hrcomp.eq0}).
For reference, the slope of the X-ray background in this energy range,
$\Gamma \sim 1.4$, corresponds to a $HR$ of $-0.22$.

\begin{figure}
\epsscale{1} 
\plotone{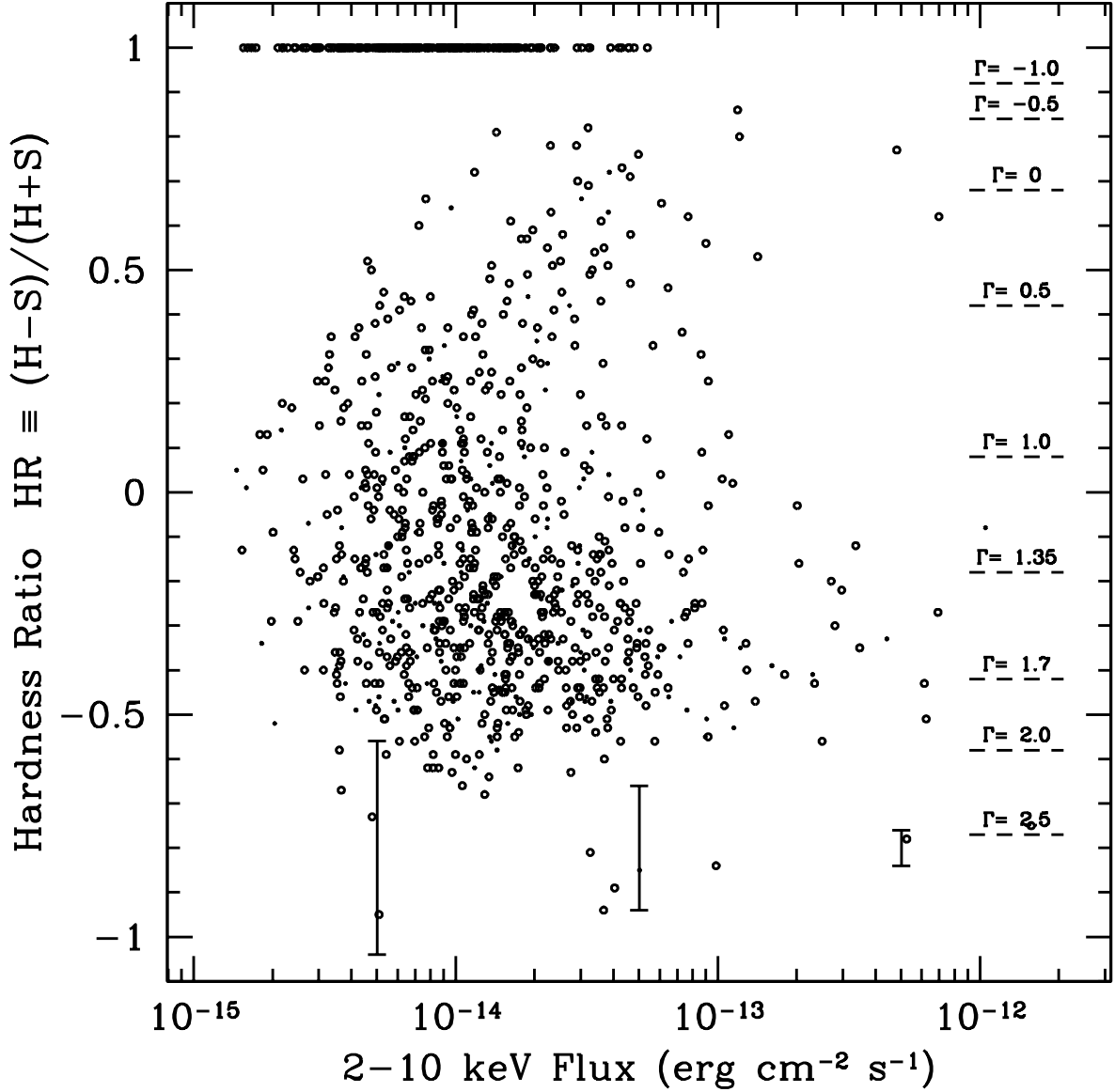} 
\caption{Hardness ratio $HR \equiv (H-S)/(H+S)$ of SEXSI 
sources as a function of hard X-ray (\hardrange) flux.  
Sources detected only in the hard X-ray
band are shown at a hardness ratio of $1$, while sources detected only
in the soft X-ray band are not shown.  Dashed horizontal lines are
power-law models with different photon indices.
The 190 sources flagged as potentially being associated with {\it
Chandra} cluster targets ($R <$ 1~Mpc; \S 3.2) are marked as dots.
The remaining 844 sources are marked as small, open circles.  Error
bars at bottom of figure show the typical uncertainties in hardness
ratio measurements at three flux levels. \label{fig:hardness}} 
\end{figure}

\begin{figure}
\epsscale{1} 
\plotone{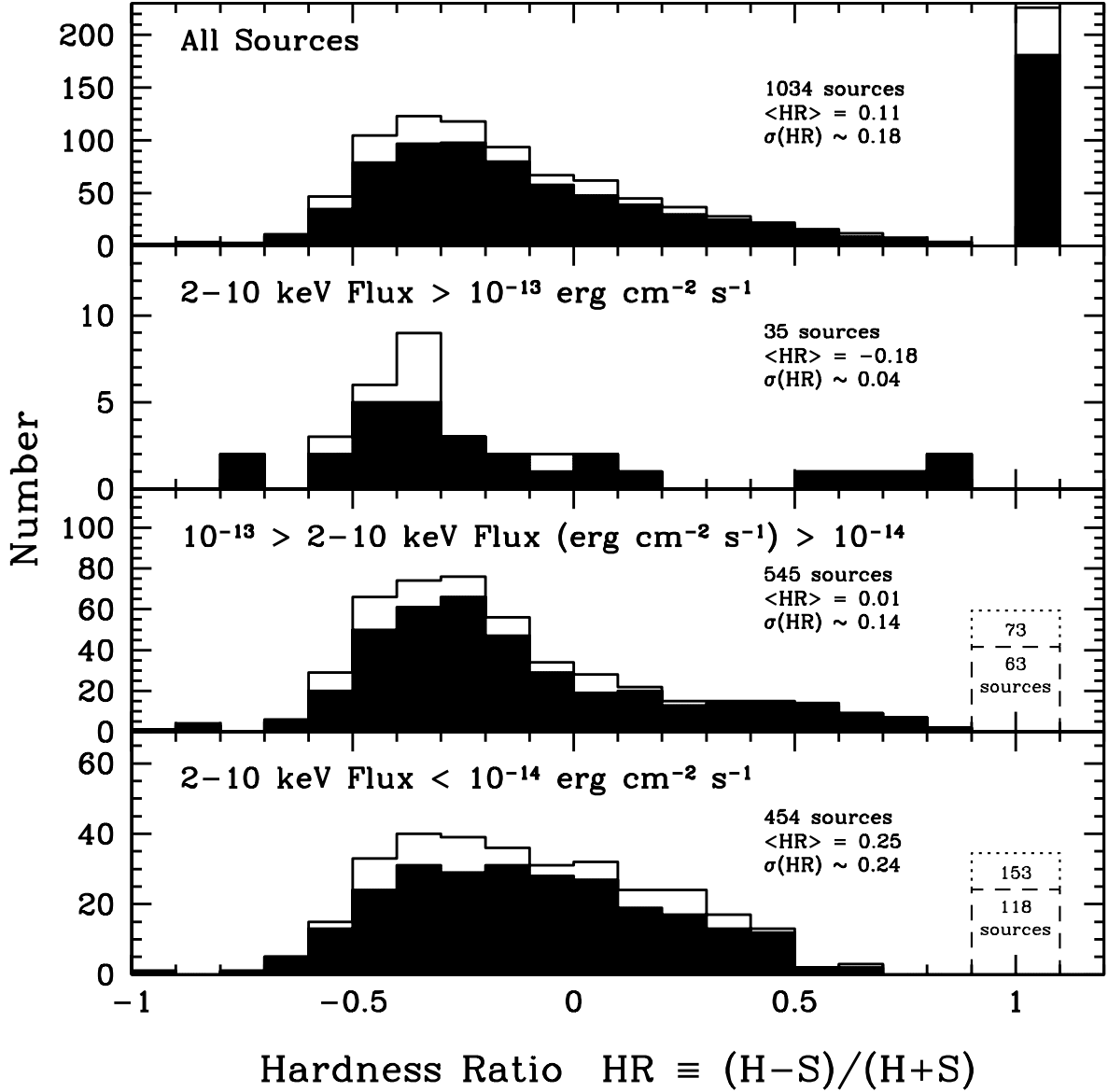} 
\caption{Histogram of hardness ratio values for the SEXSI 2$-$10~keV selected
sample of X-ray sources.  In the top panel, the open histogram shows
all 1034 sources from our survey, and the solid histogram shows the
subset of 844 sources which were not flagged as being potentially
associated with the {\it Chandra} cluster targets ($R<1$ Mpc). Note
that when statistical uncertainties are considered, the sources in the
$HR=1$ peak will partially fill in the high $HR$ end of the histogram
(see Figure~\ref{fig:hrcomp.eq0}). The lower three panels show the
data from the top panel split into three flux ranges. Sources
undetected in the soft band are indicated in text for the bottom two
panels.  Typical uncertainties in hardness ratio measurements for
individual sources in each flux bin are indicated as $\sigma(HR)$.
\label{fig:hardhist}}
\end{figure}

Figure~\ref{fig:hardness} presents the $HR$ for SEXSI's 1034 sources
as a function of hard-band flux.  The top panel of
Figure~\ref{fig:hardhist} shows these same sources in an $HR$
histogram. The lower three panels of Figure~\ref{fig:hardhist} show
the hardness ratio histogram broken into three flux ranges. The upper
right corner of each panel indicates the number of sources and average
$HR$ for each subsample. The entire sample has an average $HR$ of
$0.108 \pm 0.006$, corresponding to $\Gamma = 0.96$.  The histograms clearly
illustrate the trend, previously noted by the Ms surveys, for higher
hardness ratios at lower fluxes.

\subsection{Distribution of hardness ratios}
\label{sec:splitpops}

The highest flux (second from the top) panel in
Figure~\ref{fig:hardhist} appears to show a bimodal distribution in
hardness ratio, with a peak centered around $HR\sim-0.4$
($\Gamma\sim1.7$), and a harder, smaller peak centered around
$HR\sim0.7$ ($\Gamma\sim-0.1$).  As the flux decreases, many
of the harder sources move into the $HR = 1$ bin, while the
center of the softer distribution shifts only slightly to
the right.  This motivates us to split the hardness distribution
at $HR = 0$, and investigate the distribution of the two
sub-populations separately.

Table~\ref{tbl:hrtest} shows the result of splitting the three
flux-selected histograms at $HR = 0$, where we present the average value
of $HR_{H}$ for the six populations.  Note we use $HR_H$ to
minimize the skew imposed by the sudden shift of sources to
$HR = 1$ imposed by the requirement of separate detection in the
soft image.  The Table shows that the means for the two populations
are relatively stable as one considers fainter fluxes, but the
fraction of sources in the $HR_H<0$ population grows (see the
last column in Table~\ref{tbl:hrtest}).

\begin{figure}
\epsscale{1} 
\plotone{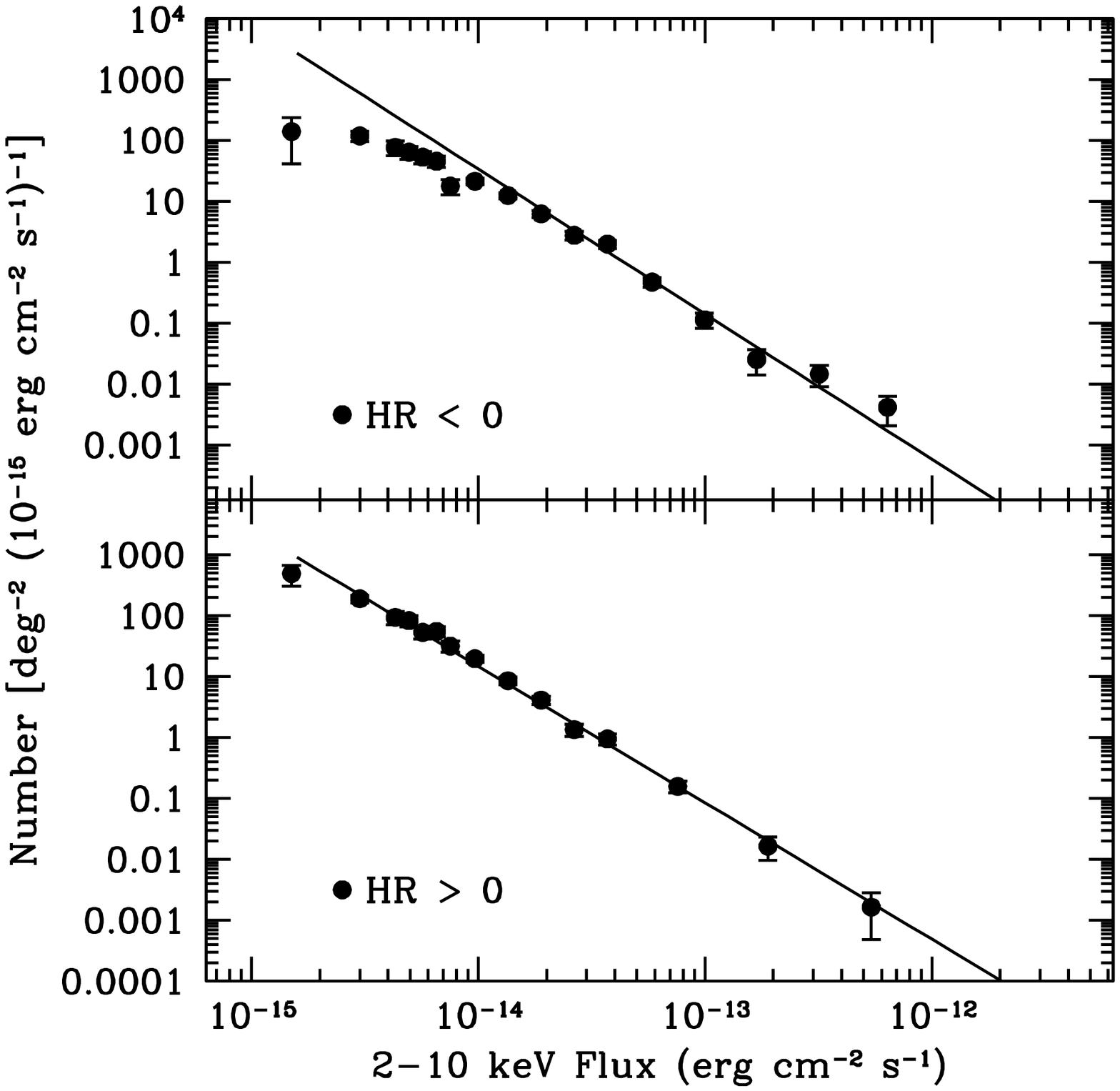} 
\caption{Differential \lognlogs~for the SEXSI sources split at 
$HR=0$. The data are plotted in units
of number deg$^{-2}$ per unit flux, where the flux has been divided
by $1 \times 10^{-15} \fluxu$. The top panel shows the $HR<0$ sources
fit from $1 \times 10^{-12}$ to $2.5 \times 10^{-14} \fluxu$. 
The bottom panel shows the $HR>0$ sources 
fit from $1 \times 10^{-12}$ to $2.5 \times 10^{-15} \fluxu$. 
The soft ($HR<0$) sources clearly turn over at $\sim$ $1 \times
10^{-14} \fluxu$ while the hard ($HR>0$) sources do not and 
are well fit by a single power law to faint fluxes.
\label{fig:splitlnn}}
\end{figure}

Figure~\ref{fig:splitlnn} shows the 
\hardrange~\lognlogs~relations for the SEXSI sources
split at $HR=0$. We have excluded the cluster fields from this
analysis to avoid bias. For the $HR<0$ plot (top panel) 
we fit the data with a single power law
at fluxes above $2.5 \times 10^{-14} \fluxu$. The best-fit curve is
parameterized by

\begin{equation}
n(S)_{HR<0.0} = (33.9 \pm_{1.5}^{1.6})(S_{\rm 2 - 10 keV}/10^{-14})^{-2.38 \pm 0.13}.
\end{equation}

The population clearly turns over at $\sim 1 \times 10^{-14} \fluxu$.
Conversely, the $HR>0.0$ population (bottom panel)
shows no break. We fit the hard data with a single power law at fluxes
all the way down to $2.5 \times 10^{-15} \fluxu$. The best-fit curve
is parameterized by
\begin{equation}
n(S)_{HR>0.0} = (14.8 \pm^{1.8}_{1.6})(S_{\rm 2 - 10 keV}/10^{-14})^{-2.24 \pm 0.05}.
\end{equation}
This curve is an excellent fit all the way down to the faint end
of our sample. Presumably the hard sources are on average at lower
redshift and thus do not exhibit the evolutionary effects likely to be
responsible for the slope break until even fainter flux levels are reached. 

\subsection{X-ray spectral comparison to previous work}

The SEXSI catalog includes only sources independently identified
in the hard band images, and so excludes those sources detected
only in the soft band.  Thus, we expect our average hardness
ratio to be significantly larger than that reported for the
deep fields, which include a large fraction of soft-only sources.
Indeed, \citet{rtg+02} analyze a stacked spectrum of the CDF-S
total sample and report an average power law index of $\Gamma=1.375$
($HR = -0.2$), much softer than our average $HR = 0.108$. Even the
faintest subsample, \hardfluxrange~$< 2 \times 10^{-15} \fluxu$, with
an average $\Gamma = 1.05$ ($HR=0.04$), appears softer than 
our entire sample.  

To make a better comparison to the CDF-S, we eliminated the soft-band
only sources from their source catalog~\citep{gzw+02}. In addition, we
translated their fluxes, which had been converted from counts using
$\Gamma=1.4$, to match ours, which assumed $\Gamma=1.5$ (a correction
of about 5\% for the hard band). We also correct for the different
spectral ranges assumed for their hard count rate measurement (2 -- 7 keV
for CDF-S compared to 2 -- 10 keV for SEXSI).  Using these converted
$HR$'s with the soft-band only sources ignored, we find that the
average $HR$ for the CDF-S sample is $0.14 \pm 0.01$, comparable to the SEXSI
$HR$ of $0.108 \pm 0.006$. Since CDF-S samples the fainter section of the
\lognlogs, their slightly higher average $HR$ is not surprising. 

To further compare the surveys, we break the CDF-S sample into three flux
ranges, as we did for our sample in Figure~\ref{fig:hardhist}.  The
CDF-S has no sources in the bright range (\hardfluxrange~$> 10^{-13}
\fluxu$).  In the medium flux range ($10^{-13}
\fluxu>$~\hardfluxrange~$ > 10^{-14} \fluxu$), we calculate the CDF-S
average $HR$ to be $-0.09 \pm 0.01$, as compared to SEXSI's average $HR$
of $0.008 \pm 0.007$. For the low flux range ($10^{-14} \fluxu>$~
\hardfluxrange ~$ > 10^{-15} \fluxu$), the 
SEXSI's average $HR$ is $0.25 \pm 0.01$ as compared to $0.13 \pm 0.02$ for CDF-S.

For each of these flux ranges we find the average $HR$'s of SEXSI and
CDF-S to be comparable, but slightly higher for SEXSI. This is likely
explained by the different survey depths and source detection
processes.  As with the SPICES reduction of the CL0848+0454
field, CDF-S detects sources in full band (0.5 -- 7 keV) images and
then extracts fluxes from the soft and hard band images regardless of
detection significance in the individual bands. For a source that is
below our threshold in the soft band we will report a flux of zero,
while CDF-S may detect positive flux.  If we compare the CDF-S $HR$'s
to our $HR_H$ values of $-0.014
\pm 0.007$ and $0.15 \pm 0.01$ for the mid- and low- flux ranges, we are
consistent with the CDF-S values of $-0.09$ and $0.16$. 

\section{Summary}

We have completed the first ``large''-area ($> 1$ deg$^2$) hard X-ray source
survey with the {\em Chandra} Observatory, and report here the X-ray 
characteristics of 1034 serendipitous sources from 27 fields detected in the 
\hardrange~band. This work represents a sample size in the critical flux interval
$1 \times 10^{-13}$ to $3 \times 10^{-15} \fluxu$ that exceeds the sum
of all previous surveys by a factor of three. We present a technique
for calculating the effective area of our survey which is fully
consistent with our source detection algorithm; combined with the
large source sample, this allows us to derive the most accurate
\lognlogs~relation yet produced for hard X-ray sources at fluxes
fainter than $10^{-13} \fluxu$. We find that the slope of the relation
is Euclidean at fluxes above $10^{-14} \fluxu$. Combining the complete
source sample with the CDF-S deep survey data indicates a break in the
\lognlogs~slope at $1.1 \times 10^{-14} \fluxu$. Calculation of
separate \lognlogs~ relations for the hard and soft portions of our
sample shows that it is the softer hard-band sources which are responsible for
this break; sources with $HR>0.0$ show no slope change down to a flux
an order of magnitude fainter, suggesting (as our spectroscopic
followup and that of the deep surveys of \citet{hbg+01} and 
\citet{trn+01} have confirmed) that the hardest sources are
predominantly a lower redshift sample. Future papers in this series
will describe our optical observations of this sample, providing
further insight into the populations of X-ray luminous objects that
comprise the X-ray background.

\acknowledgements

We are deeply indebted to the following {\em Chandra} Guest
Investigators for their willingness to allow us immediate access to
their data for the purposes of this statistical investigation of
serendipitous sources: Dr. Jan Vrtilek, Dr. Crystal Martin, and
Prof. Q. Daniel Wang. We also thank Dr. Adam Stanford and his
collaborators for their enlightened approach to the deepest pointing
included in our survey that waived all proprietary rights to these
data and Dr. Leon van Speybroeck for contributing three of his GTO
cluster fields.  Elise Laird and Alan Diercks assisted greatly in the
construction of our optical data reduction pipeline with useful code
and helpful advice. James Chakan assisted with the X-ray data
reduction. This work has been supported by NASA NAG5-6035(DJH), as
well as by a small {\em Chandra} archival grant. The work of DS was
carried out at the Jet Propulsion Laboratory, California Institute of
Technology, under a contract with NASA.

\section*{APPENDIX}
\label{sec:appendix}

In Table~\ref{tbl:softcatalog} we present a source catalog of 879
soft-band serendipitous X-ray sources which lack a statistically significant 
hard-band counterpart. These sources are excluded from the main
SEXSI catalog (Table~\ref{tbl:catalog}) since the strength of SEXSI, 
and thus our primary scientific interest, lies in the study of \hardrange~source 
populations. 

These soft sources, detected and analyzed as described in 
Section~\ref{sec:reduction}, are designated CXOSEXSI (our IAU-registered name)
followed by standard truncated source coordinates. The source
positions (equinox J2000.0) are those derived from the soft-band X-ray
images. We include only sources
detected with a chance coincidence probability of $<10^{-6}$ in the
soft band. The angular distance of the source position from the
telescope axis is given in column 4.  Columns 5 and 6 list the
background-subtracted counts for each source within the specified
aperture derived from the 0.5 -- 2.1 keV image, followed by the
estimated background counts in that same aperture. Column 7 gives an
estimate of the signal-to-noise ratio (SNR) of the detection (see 
Section~\ref{sec:catalog} for details).  
Again, it should be emphasized that these values are {\it not} a
measure of source significance (which is $P<10^{-6}$ in all
cases) but is a measure of the uncertainty in the source flux
estimates. Column 8 shows the unabsorbed soft band flux (with units of $10^{-15} \fluxu$), 
corrected for source counts falling outside the aperture and translated to the
standard \softrange~band assuming a power law photon spectral index of $\Gamma
=1.5$ and a Galactic absorbing column density appropriate to the field
(see Table~\ref{tbl:fields}). 

This soft-band only catalog does have the target sources carefully 
eliminated for point sources and nearby galaxies, as described for 
the main SEXSI catalog in Section~\ref{sec:srcdeletions}. This led 
to the removal of 86 sources from this catalog.
However, the sources within $\sim 1$ Mpc of target galaxy cluster 
centroids are not flagged, as was done with the hard sources in 
Table~\ref{tbl:catalog}. In addition, there has been no attempt 
to search for extended sources.

\pagebreak[4]

\begin{deluxetable}{ccccccccc}
\tablecaption{Comparison of Sources Detected in 2 -- 10 keV Hard X-Ray Surveys.}
\tablecolumns{9}
\tablewidth{0pc}
\tablehead{\colhead{} & \multicolumn{6}{c}{PUBLISHED SURVEYS}  & \colhead{} & \colhead{} \\ 
\cline{2-8}
\colhead{log Flux Range} & \colhead{{\em ASCA}\tablenotemark{a}} & \colhead{{\em SAX}\tablenotemark{b}} & \colhead{SSA13\tablenotemark{c}} & \colhead{CDF-N\tablenotemark{d}} & \colhead{CDF-S\tablenotemark{e}} & \colhead{Lynx\tablenotemark{f}} & \colhead{TOTAL} & \colhead{SEXSI}}
\startdata
--14.5 $-$ --15.0 &  0 &    0  &   8  &   106  &    92   &    49  &   255  &     55 \\
--14.0 $-$ --14.5 &  0 &    0  &  18  &    56  &    66   &    45  &   185  &    400 \\
--13.5 $-$ --14.0 &  0 &    0  &   6  &    23  &    21   &    12  &    62  &    399 \\
--13.0 $-$ --13.5 &  2 &   17  &   1  &     5  &     4   &     4  &    33  &    145 \\
--12.5 $-$ --13.0 & 51 &   89  &   0  &     1  &     0   &     1  &   142  &     24 \\
--12.0 $-$ --12.5 & 35 &   61  &   0  &     0  &     0   &     0  &    96  &      9 \\
--11.5 $-$ --12.0 &  5 &    9  &   0  &     0  &     0   &     0  &    14  &      2 \\
--11.0 $-$ --11.5 &  1 &    1  &   0  &     0  &     0   &     0  &     2  &      0 \\ 
\cline{1-9} 
TOTALS     & 94 &  177  &  33  &   191  &   183   &   111  &   789  &   1034 \\
\enddata
\label{tbl:surveycomp}
\footnotesize
\tablenotetext{a}{For each survey, we provide the primary reference(s), the satellite and X-ray instrument used,
the spectral assumptions adopted, and the factor by which we multiplied the tabulated fluxes to
bring them into conformity with the energy band and spectral parameters adopted in our study.
For {\em ASCA}, see \citet{cdm+98}; GIS2-selected; $\Gamma = 1.7,$ actual $N_{H}$ ~($N_H \sim 3
\times 10^{20}$ cm$^{-2}$); factor $=\times 1.06$. In addition, see \citet{aoy+00}; SIS-selected;
best PL model and $N_{H} (N_{H}\sim 3 \times 10^{20}$) from SIS + GIS fit;
factor based on individual spectral indices ($=\times 0.52 - 1.36$).}
\tablenotetext{b}{For {\em SAX}, see \citet{gpf+00}; MECS-selected; $\Gamma = 1.7,$ actual $N_{H} (\sim 3
\times 10^{20}$ cm$^{-2}$); factor $=\times 0.959$.}
\tablenotetext{c}{For SSA13, see \citet{mcb+00}; ACIS-S-selected; $\Gamma = 1.2$, actual $N_{H} = 1.4
\times 10^{20}$ cm$^{-2}$; factor $=\times 0.986$ and, for chip 3 only, $=\times 0.948$}
\tablenotetext{d}{For the CDF-N, see \citet{bah+01}; ACIS-I-selected; hardness-ratio-derived 
spectral slopes, $N_{H} = 1.6 \times 10^{20}$ cm$^{-2}$ {\it not} included;
 \citet{cgb+02} claim the mean flux is increased by 13\% from assuming 
$\Gamma=1.2$, so factor $=\times 1.293$ (to get \hardrange~intrinsic flux) $\times 0.885$ to get all sources to $\Gamma = 1.2$
$\times 0.9345$ to get $\Gamma = 1.5$, so final factor $= \times 1.069$}
\tablenotetext{e}{For the CDF-S, see \citet{gzw+02}; ACIS-I-selected; $\Gamma = 1.375, 
N_{H} = 0.8 \times 10^{20}$ cm$^{-2}$; factor $=\times 0.932$}
\tablenotetext{f}{For the Lynx field, see \citet{sts+02}; ACIS-I-selected; $\Gamma = 1.4, N_{H} =
2 \times 10^{20}$ cm$^{-2}$; factor $=\times 1.004$}
\tablecomments{\citet{utt+01} have recently published a catalog of \hardrange~X-ray sources from
the {\em ASCA} database which contains 1343 sources. Of these, 4 have a detection significance 
in the \hardrange~band of $\gsim 3.0 \sigma$ and $1 \times 10^{-14} \fluxu 
<$~\hardfluxrange~$< 3\times 10^{-14} \fluxu$,
while 112 entries lie in the range $3 \times 10^{-14} \fluxu <$~\hardfluxrange~$
< 1\times 10^{-13} \fluxu$. However,
the effective area covered by this survey as a function of flux and the
\lognlogs~curves have not been presented, so we have not included these sources
in the above table.}
\end{deluxetable}

\begin{deluxetable}{lllccccc}
\tabletypesize{\footnotesize}
\tablecaption{Chandra Observations.}
\tablecolumns{8}
\tablewidth{0pc}
\tablehead{\multicolumn{3}{c}{Target} & \colhead{ } &\colhead{ } &\colhead{ } &\colhead{ } &\colhead{ } \\  \\
\cline{1-3}
\colhead{ } &\colhead{ } &\colhead{ } & \colhead{RA} & \colhead{DEC} & \colhead{N$_{H}$}& \colhead{exp.} & \colhead{ACIS chips}\\
Name & Type& $z$ or $cz$ & \colhead{(J2000)} & \colhead{(J2000)} &\colhead{[$10^{20}$ cm$^{-2}$]} & \colhead{[ks]} & \colhead{ }}
\startdata
NGC 891    & edge-on spiral        & 528 km s$^{-1}$   & 02 22 33 &  +42 20 57  & 6.7&  51 & S 235-8 \\ 
AWM 7      & galaxy cluster        & 0.017      & 02 54 28 &  +41 34 47  & 9.2&  48 & I:0-367 \\ 
XRF 011130 & X-ray flash afterglow & \nodata      & 03 05 28 &  +03 49 59  & 9.3&  30 & I:0-3   \\ 
NGC 1569   & spiral galaxy         & $-$104 km s$^{-1}$  & 04 30 49 &  +64 50 54  &23.8&  97 & S:2357 \\ 
3C 123     & galaxy cluster        & 0.218      & 04 37 55 &  +29 40 14  &19.0&  47 & S:235-8 \\ 
CL 0442+2200   & galaxy cluster        & 1.11       & 04 42 26 &  +02 00 07  & 9.5&  44 & I:0-3  \\
CL 0848+4454   & galaxy cluster        & 1.27       & 08 48 32 &  +44 53 56  & 2.8& 186 & I:0-367 \\ 
RX J0910   & galaxy cluster        & 1.11       & 09 10 39 &  +54 19 57  & 1.9& 171 & I:0-36   \\ 
1156+295   & blazar                & 0.729      & 11 59 32 &  +29 14 44  & 1.7&  49 & I:0-3   \\ 
NGC 4244   & edge-on spiral        & 244 km s$^{-1}$   & 12 17 30 &  +37 48 32  & 1.8&  49 & S:235-8 \\ 
NGC 4631   & edge-on disk galaxy   & 606 km s$^{-1}$   & 12 42 07 &  +32 32 30  & 1.2&  59 & S:235-8 \\ 
HCG 62     & compact group         & 4200 km s$^{-1}$  & 12 53 08 & $-$09 13 27  & 2.9&  49 & S:6-8   \\ 
RX J1317   & galaxy cluster        & 0.805      & 13 17 12 &  +29 11 17  & 1.1& 111 & I:0-36  \\ 
BD 1338    & galaxy cluster        & 0.640      & 13 38 25 &  +29 31 05  & 1.1&  38 & I:0-36  \\ 
RX J1350   & galaxy cluster        & 0.804      & 13 50 55 &  +60 05 09  & 1.8&  58 & I:0-36  \\ 
3C 295     & galaxy cluster        & 0.46       & 14 11 20 &  +52 12 21  & 1.3&  23 & S:236-8 \\ 
GRB 010222 & GRB afterglow         & 1.477      & 14 52 12 &  +43 01 44  & 1.7&  18 & S:236-8 \\ 
QSO 1508   & quasar                & 4.301      & 15 09 58 &  +57 02 32  & 1.4&  89 & S:2367  \\ 
MKW 3S     & galaxy cluster        & 0.045      & 15 21 52 &  +07 42 32  & 2.9&  57 & I:0-368 \\ 
MS 1621    & galaxy cluster        & 0.4281     & 16 23 36 &  +26 33 50  & 3.6&  30 & I:0-36  \\ 
GRB 000926 & GRB afterglow         & 2.038      & 17 04 10 &  +51 47 11  & 2.7&  32 & S:236-8 \\ 
RX J1716   & galaxy cluster        & 0.81       & 17 16 52 &  +67 08 31  & 3.8&  52 & I:0-36  \\ 
NGC 6543   & planetary nebula      &  0         & 17 58 29 &  +66 38 29  & 4.3&  46 & S:5-9   \\ 
XRF 011030 & X-ray flash afterglow & \nodata      & 20 43 32 &  +77 16 43  & 9.5&  47 & S:2367 \\ 
MS 2053    & galaxy cluster        & 0.583      & 20 56 22 &$-$04 37 44  & 5.0&  44 & I:0-36  \\ 
RX J2247   & galaxy cluster        & 0.18       & 22 47 29 &  +03 37 13  & 5.0&  49 & I:0-36  \\ 
Q2345      & quasar pair           & 2.15       & 23 48 20 &  +00 57 21  & 3.6&  74 & S:2678  \\
\cline{1-8}
TOTAL & & & & & & 1648 & 134 \\
\enddata
\label{tbl:fields}
\end{deluxetable}

\begin{deluxetable}{ll}
\tablecolumns{2}
\tablecaption{Percent of PSF half-width as a function of off-axis 
angle used to extract hard counts (the value for soft count extraction
is obtained by adding 0.06\% to the hard counts \% of PSF half-width value). 
These values optimize the signal-to-noise of each detection (see \S\ref{sec:reduction}).}
\tablewidth{0pc}
\tablehead{\colhead{OAA [\arcmin]} & \colhead{\% of PSF HW}}
\startdata
$0-3$ & $90.0$ \\
$3-5$ & $87.5$ \\
$5-10$ & $85.0$ \\
$>10$ & $80.0$ \\
\enddata
\label{tbl:psfoaa}
\end{deluxetable}

\pagebreak[4]

\begin{deluxetable}{lccrrrrrrrrrr}
\tablecaption{Source Catalog.}
\tabletypesize{\footnotesize}
\rotate
\tablecolumns{13}
\tablewidth{0pc}
\tablehead{\colhead{CXOSEXSI\_}  & \colhead{RA} & \colhead{DEC} & \colhead{OAA} & \multicolumn{4}{c}{\underline{~~~~~~~~~~~~~Hard Band~~~~~~~~~~~~~}}&\multicolumn{4}{c}{\underline{~~~~~~~~~~~~~Soft Band~~~~~~~~~~~~~}}& \colhead{HR} \\ 
\colhead{ } & \colhead{(J2000)}& \colhead{(J2000)} & \colhead{[\arcmin]}& \colhead{Cts} & \colhead{Bkg} & \colhead{SNR} & \colhead{Flux\tablenotemark{a}}& \colhead{Cts} & \colhead{Bkg} & \colhead{SNR} & \colhead{Flux\tablenotemark{a}}& \colhead{ } \\
\colhead{(1)} & \colhead{(2)} & \colhead{(3)} & \colhead{(4)} & \colhead{(5)} & \colhead{(6)} & \colhead{(7)} & \colhead{(8)} & \colhead{(9)} & \colhead{(10)} & \colhead{(11)} & \colhead{(12)} & \colhead{(13)}}
\startdata
J022142.6+422654                   &  02 21 42.67 &     42 26 54.1 &  9.49 &  16.30 &  5.70 &  2.83 &  23.10 &  48.50 &  6.50 &  5.73 &  10.30 & -0.33 \\
J022143.6+421631                   &  02 21 43.64 &     42 16 31.6 &  8.33 &  98.53 &  4.47 &  8.81 &  74.80 & 244.27 &  3.73 & 14.56 &  29.70 & -0.28 \\
J022151.6+422319                   &  02 21 51.68 &     42 23 19.3 &  6.17 &   9.13 &  1.87 &  2.06 &   6.55 &  17.67 &  1.33 &  3.25 &   2.02 & -0.16 \\
J022205.0+422338                   &  02 22 05.00 &     42 23 38.3 &  4.24 &  13.24 &  0.76 &  2.74 &   8.73 &   0.00 &  0.00 &  0.00 &   0.00 &  1.00 \\
J022205.1+422213                   &  02 22 05.13 &     42 22 13.3 &  3.45 &  10.55 &  0.45 &  2.38 &   6.82 &  12.61 &  0.39 &  2.68 &   1.33 &  0.07 \\
J022207.1+422918                   &  02 22 07.11 &     42 29 18.8 &  8.93 &  25.10 &  4.90 &  3.84 &  18.90 &  64.86 &  4.14 &  6.94 &   8.46 & -0.33 \\
J022210.0+422956                   &  02 22 10.00 &     42 29 56.3 &  9.38 &  22.38 &  5.62 &  3.52 &  17.30 & 127.12 &  4.88 & 10.15 &  16.70 & -0.62 \\
J022210.8+422016                   &  02 22 10.85 &     42 20 16.7 &  2.17 &   5.52 &  0.48 &  1.53 &   3.63 &  19.42 &  0.58 &  3.50 &   2.17 & -0.46 \\
J022211.7+421910                   &  02 22 11.71 &     42 19 10.7 &  2.55 &  10.46 &  0.54 &  2.36 &   6.87 &  10.40 &  0.60 &  2.35 &   1.15 &  0.14 \\
J022215.0+422341                   &  02 22 15.04 &     42 23 41.6 &  3.15 &  23.60 &  1.40 &  3.88 &  15.40 & 122.76 &  1.24 & 10.09 &   8.13 & -0.40 \\
J022215.1+422045                   &  02 22 15.11 &     42 20 45.1 &  1.32 &  72.69 &  2.31 &  7.49 &  42.90 & 378.26 &  3.74 & 18.39 &  23.70 & -0.42 \\
J022215.5+421842                   &  02 22 15.55 &     42 18 42.3 &  2.46 &  11.33 &  2.67 &  2.34 &   6.79 &  13.18 &  3.82 &  2.53 &   0.85 &  0.28 \\
J022219.3+422052                   &  02 22 19.32 &     42 20 52.2 &  0.54 &  10.89 &  2.11 &  2.31 &   6.38 &   0.00 &  0.00 &  0.00 &   0.00 &  1.00 \\
J022224.3+422139                   &  02 22 24.37 &     42 21 39.0 &  0.91 &  98.89 &  2.11 &  8.92 &  64.60 & 529.79 &  4.21 & 21.96 &  37.20 & -0.44 \\
J022225.2+422451                   &  02 22 25.25 &     42 24 51.5 &  4.07 & 196.06 &  1.94 & 12.99 & 128.00 & 852.89 &  3.11 & 28.18 &  57.80 & -0.34 \\
J022226.5+422155                   &  02 22 26.55 &     42 21 55.0 &  1.35 &  10.90 &  2.10 &  2.32 &   6.48 &  47.75 &  4.25 &  5.78 &   3.03 & -0.35 \\
J022232.5+423015                   &  02 22 32.53 &     42 30 15.2 &  9.61 &  59.97 &  7.03 &  6.50 &  53.60 &  61.20 &  4.80 &  6.67 &   9.47 &  0.12 \\
J022236.3+421730                   &  02 22 36.37 &     42 17 30.8 &  4.23 &  19.39 &  3.61 &  3.30 &  12.00 &  40.32 &  2.68 &  5.30 &   2.64 &  0.01 \\
J022236.8+422858                   &  02 22 36.80 &     42 28 58.3 &  8.57 &  17.32 &  4.68 &  3.00 &  12.90 & 115.64 &  3.36 &  9.68 &  15.10 & -0.68 \\
J022259.1+422434                   &  02 22 59.10 &     42 24 34.2 &  7.77 &  42.98 &  4.02 &  5.43 &  32.50 &  17.51 &  4.49 &  3.03 &   2.49 &  0.49 \\
J022334.0+422212                   &  02 23 34.05 &     42 22 12.2 & 13.34 &  40.98 & 25.02 &  4.47 &  37.50 &  37.78 & 26.22 &  4.18 &   6.21 &  0.15 \\
J025325.9+413941 \tablenotemark{b} &  02 53 25.98 &     41 39 41.2 & 13.76 &  48.27 & 52.73 &  4.35 &  50.30 & 836.51 & 120.5 & 26.18 & 141.00 & -0.85 \\
J025333.7+413928 \tablenotemark{b} &  02 53 33.74 &     41 39 28.0 & 12.31 &  62.72 & 45.28 &  5.49 &  62.50 & 180.10 & 122.9 &  9.77 &  29.30 & -0.35 \\
J025340.4+413019 \tablenotemark{b} &  02 53 40.44 &     41 30 19.8 & 14.60 &  87.55 & 158.4 &  5.24 &  89.60 & 647.21 & 524.7 & 18.36 &  69.40 & -0.55 \\
J025400.3+414006 \tablenotemark{b} &  02 54 00.32 &     41 40 06.4 &  7.34 & 286.41 &  9.59 & 15.71 & 230.00 & 804.34 & 19.66 & 27.07 & 122.00 & -0.41 \\
\enddata
\tablenotetext{a} {Fluxes are presented in units of $10^{-15} \fluxu$.}
\tablenotetext{b} {Source falls within excluded area less than 1Mpc from cluster center. Source not used for \lognlogs~calculation.} 
\label{tbl:catalog}
\end{deluxetable}

\begin{deluxetable}{lccccc}
\tablecolumns{6}
\tablecaption{$HR_H$ averages for sources with $HR_H<0$ and $HR_H>0$ (see \S\ref{sec:splitpops}).}
\tablewidth{0pc}
\tablehead{\colhead{Flux Range} & \multicolumn{2}{c}{\underline{~~~~~~~$HR_{H} < 0.0$~~~~~~~}} & \multicolumn{3}{c}{\underline{~~~~~~~~~~~~~$HR_{H} > 0.0$~~~~~~~~~~~~~}} \\
\colhead{$\fluxu$} & \colhead{$\langle HR_{H} \rangle$} & \colhead{\# Srcs} & \colhead{$\langle HR_{H} \rangle$} & \colhead{\# Srcs} 
& \colhead{\% of Srcs}}
\startdata
$10^{-12} - 10^{-13}$ & $-0.38 \pm 0.01$& 26 &  $0.42  \pm 0.02$& 9  &  26\\
$10^{-13} - 10^{-14}$ & $-0.31 \pm 0.01$& 344 & $0.49 \pm 0.01$& 201 & 37\\
$10^{-14} - 10^{-15}$ & $-0.29 \pm 0.02$& 201 & $0.50 \pm 0.02$& 253 & 56\\
\enddata
\label{tbl:hrtest}
\end{deluxetable}

\pagebreak[4]

\pagebreak[4]

\begin{deluxetable}{lccrrrrr}
\tablecaption{Soft-band Only Source Catalog.}
\tablecolumns{8}
\tablewidth{0pc}
\tablehead{\colhead{CXOSEXSI\_}  & \colhead{RA} & \colhead{DEC} & \colhead{OAA} & \multicolumn{4}{c}{\underline{~~~~~~~~~~~~~Soft Band~~~~~~~~~~~~~}}\\ 
\colhead{ } & \colhead{(J2000)}& \colhead{(J2000)} & \colhead{[\arcmin]}& \colhead{Cts} & \colhead{Bkg} & \colhead{SNR} & \colhead{Flux\tablenotemark{a}}\\
\colhead{(1)} & \colhead{(2)} & \colhead{(3)} & \colhead{(4)} & \colhead{(5)} & \colhead{(6)} & \colhead{(7)} & \colhead{(8)}}
\startdata
J022045.8+421954                      &  02 20 45.80 &     42 19 54.2 & 17.84 &  387.68 & 449.3 & 12.95 &  37.90 \\
J022054.2+421724                      &  02 20 54.22 &     42 17 24.3 & 16.62 &  421.68 & 482.3 & 13.57 &  40.60 \\
J022101.0+422042                      &  02 21 01.08 &     42 20 42.4 & 14.99 &  123.61 & 199.3 &  6.51 &  11.20 \\
J022108.5+422008                      &  02 21 08.52 &     42 20 08.2 & 13.64 &  109.91 & 206.0 &  5.85 &   9.88 \\
J022111.1+421704                      &  02 21 11.11 &     42 17 04.1 & 13.67 &   98.28 & 199.7 &  5.38 &   8.69 \\
J022113.3+421842                      &  02 21 13.35 &     42 18 42.3 & 12.90 &  197.24 & 163.7 &  9.85 &  17.40 \\
J022122.8+421725                      &  02 21 22.84 &     42 17 25.1 & 11.49 &  190.08 & 87.92 & 10.74 &  16.10 \\
J022128.4+421826                      &  02 21 28.41 &     42 18 26.2 & 10.23 &   54.50 &  7.50 &  6.11 &   8.94 \\
J022131.1+422146                      &  02 21 31.15 &     42 21 46.5 &  9.48 &  1800.8 &  8.19 & 41.36 & 222.00 \\
J022131.2+422144                      &  02 21 31.24 &     42 21 44.0 &  9.46 &  1659.9 &  8.10 & 39.66 & 204.00 \\
J022131.5+423103                      &  02 21 31.52 &     42 31 03.3 & 13.86 &   45.62 & 28.38 &  4.73 &   6.99 \\
J022131.5+422148                      &  02 21 31.58 &     42 21 48.3 &  9.41 & 1649.00 &  7.92 & 39.53 & 203.00 \\
J022131.6+422144                      &  02 21 31.61 &     42 21 44.1 &  9.39 & 1574.00 &  7.94 & 38.60 & 194.00 \\
J022136.1+422730                      &  02 21 36.11 &     42 27 30.7 & 10.82 &   19.42 &  9.58 &  3.01 &   3.00 \\
J022140.9+422050                      &  02 21 40.95 &     42 20 50.2 &  7.62 &   20.86 &  3.14 &  3.49 &   2.48 \\
J022142.1+421947                      &  02 21 42.18 &     42 19 47.4 &  7.47 &   26.10 &  2.90 &  4.04 &   3.10 \\
J022144.2+423019                      &  02 21 44.29 &     42 30 19.4 & 11.79 &   25.28 & 13.72 &  3.46 &   3.70 \\
J022153.5+423026                      &  02 21 53.50 &     42 30 26.1 & 10.97 &   25.00 &  8.00 &  3.67 &   3.59 \\
J022155.1+421804                      &  02 21 55.13 &     42 18 04.3 &  5.72 &    7.90 &  1.10 &  1.92 &   0.91 \\
J022155.5+421749                      &  02 21 55.56 &     42 17 49.8 &  5.77 &    7.84 &  1.16 &  1.90 &   0.90 \\
J022202.6+421637                      &  02 22 02.69 &     42 16 37.7 &  5.54 &   14.00 &  1.00 &  2.82 &   1.79 \\
J022205.9+421652                      &  02 22 05.90 &     42 16 52.3 &  4.98 &   27.05 &  0.95 &  4.25 &   2.99 \\
J022221.6+422348                      &  02 22 21.60 &     42 23 48.3 &  2.98 &   11.96 &  3.04 &  2.41 &   0.82 \\
J022225.6+423526                      &  02 22 25.61 &     42 35 26.7 & 14.63 &  115.19 & 32.81 &  8.73 &  18.10 \\
J022225.7+422847                      &  02 22 25.72 &     42 28 47.1 &  7.98 &   16.38 &  2.62 &  3.01 &   2.13 \\
J022227.1+422336                      &  02 22 27.18 &     42 23 36.6 &  2.93 &   34.28 &  3.72 &  4.75 &   2.21 \\
J022227.5+422108                      &  02 22 27.59 &     42 21 08.9 &  1.04 &  168.54 &  5.46 & 11.85 &  10.80 \\
J022229.3+422852                      &  02 22 29.30 &     42 28 52.7 &  8.15 &   16.07 &  2.93 &  2.95 &   2.09 \\
\enddata
\label{tbl:softcatalog}
\tablenotetext{a} {Fluxes are presented in units of $10^{-15} \fluxu$.}
\end{deluxetable}

\end{document}